\documentclass[%
 aip,
  pop, % Physics of Plasmas substyle of AIP, no titles in citations
 amsmath,amssymb,
  reprint,%
]{revtex4-1}
\usepackage{CJK}

\usepackage{ulem}
\usepackage[T1]{fontenc}
\usepackage{calligra}

\usepackage{graphicx}% Include figure files
\usepackage{dcolumn}% Align table columns on decimal point
\usepackage{bm}% bold math

\usepackage{times}
\usepackage{verbatim}
\usepackage{graphicx}
\usepackage{graphics,epsfig}

\usepackage{theorem}
\usepackage{makeidx}
\usepackage{amsmath}
\usepackage{epic}
\usepackage{amscd}

\usepackage{bbm}
\usepackage{xy}
\usepackage{amssymb}
\usepackage{epsfig}
\def \vth {\mbox{$v_{\mbox{\scriptsize{th}}}$}}

\newcommand{\Perp}{{\mbox{$\scriptscriptstyle \perp$}}}
\newcommand{\Par}{{\mbox{$\scriptscriptstyle \|$}}}

\newcommand{\ignore}[1]{}  % a command which does nothing.
\newcommand{\sumkp}{\tilde{\sum}_{\bf k}}
\def \phi {\mbox{$\varphi$}}

\providecommand\bnabla{\boldsymbol{\nabla}}

\newsavebox{\astrutbox}
\sbox{\astrutbox}{\rule[-5pt]{0pt}{20pt}}

\newcommand{\bea}{\begin{eqnarray}}
\newcommand{\eea}{\end{eqnarray}}
\newcommand{\beq}{\begin{equation}}
\newcommand{\eeq}{\end{equation}}

\begin{document}
\begin{CJK*}{GB}{gbsn}

\preprint{AIP/123-QED}

%\title[manuscript for PoF]{Statistical mechanics of truncated gyrokinetic collisionless plasma}% Force line breaks with \\

\title[Gyrokinetic statistical absolute equilibrium and turbulence]{Gyrokinetic statistical absolute equilibrium and turbulence}

% \thanks{Footnote to title of article.}

\author{Jian-Zhou Zhu (朱建州)}
 \altaffiliation[Also at ]{Princeton Plasma Physics Laboratory.}%Lines break automatically or can be forced with \\

% GWH: add line break to match the PoP formatting of addresses:
\affiliation{%\mbox{
Center for Multiscale Plasma Dynamics, University of
  Maryland, College Park, \\
Maryland, 20742-3511, USA}%}

%\\This line break forced with

\author{Gregory W. Hammett (格里高里$\cdot$伟恩$\cdot$哈米特)}
% \homepage{http://www.Second.institution.edu/~Charlie.Author.}
\affiliation{
%\mbox{
Princeton Plasma Physics Laboratory, Princeton University,
P.O. Box 451,\\
Princeton, New Jersey 08543, USA}%}

\date{Received 23 June 2010 and accepted 19 October 2010 by {\it Physics of Plasmas}}
%\date{November 4, 2010, to be published in {\it Physics of Plasmas}}

\begin{abstract}
A paradigm based on the absolute equilibrium of Galerkin-truncated
inviscid systems to aid in understanding turbulence [T.-D. Lee, ``On
some statistical properties of hydrodynamical and magnetohydrodynamical fields,'' Q. Appl. Math. 10, 69 (1952)] is taken
to study gyrokinetic plasma turbulence: A finite set of Fourier
modes of the collisionless gyrokinetic equations are kept and the
statistical equilibria are calculated; possible implications for plasma
turbulence in various situations are discussed. For the case of two spatial and one velocity dimension, in the
calculation with discretization also of velocity $v$ with $N$ grid
points (where $N+1$
quantities are conserved, corresponding to an energy invariant and $N$ entropy-related invariants), the negative temperature states, corresponding to the condensation of
the
generalized energy into the lowest modes, are found. This indicates a generic feature of inverse energy
cascade.
Comparisons are made with some classical results, such as those of
Charney-Hasegawa-Mima in the cold-ion limit. There is a universal shape for statistical equilibrium of
gyrokinetics in three spatial and two velocity dimensions with just one
conserved quantity.
Possible physical relevance to turbulence, such as ITG zonal flows, and
to a critical balance hypothesis are also discussed.

%
%Valid PACS numbers may be entered using the \verb+\pacs{#1}+ command.
\end{abstract}

% \pacs{Valid PACS appear here}% PACS, the Physics and Astronomy
% \keywords{Suggested keywords}%Use showkeys class option if keyword

\maketitle
\end{CJK*}

\thispagestyle{plain}

\section{\label{sec:Introduction}Introduction}
Plasma dynamics encompasses a hierarchy of scales with distinct physical
processes. At scales much larger than the mean free path and gyroradius,
and time scales much larger than the collision time and gyroperiod,
the magnetohydrodynamics (MHD) model is good
(and is often quite useful over a wider range of collisionality,
particularly for phenomena where the parallel kinetic dynamics are not
important); while, in the opposite limit of high frequencies and small
scales, a complete kinetic description with the Boltzmann or Vlasov
equation is necessary.  In between, for frequencies well below the ion
cyclotron frequency but that may still involve scales comparable to the
gyroradius, a detail of the particle helical motion around the
field line, the cyclotron angle, may be averaged out, resulting in a
reduced system called
gyrokinetics.\cite{Frieman-Chen,wwLee83,BH07review,Sch09Review,Howes06,Krommes10}
With one dimension
(the cyclotron angle) and the fast time scales associated with that
dimension excluded,
gyrokinetics helps the tractability of turbulent kinetic cascades of
plasma turbulence numerically and analytically.

In this contribution, we will present the equilibrium statistical
mechanics of the Fourier Galerkin truncated gyrokinetic system and discuss the
possible implications for plasma turbulence.

Equilibrium-statistical-mechanics approaches to explore turbulence have
long been attempted to identify the flows or to provide some relevant
solutions to track the mechanisms of fluid turbulent motions, which have
been very illuminating and promising, if not completely successful.
\cite{UrielBook,GregSreeniReview} One simple but efficient strategy,
initiated by Lee,\cite{Lee1952} is calculating the Gibbs statistics of
the Galerkin-truncated system:
The flow of the Euler equation in phase space is incompressible (where
the coordinate axes $\sigma_i(\textbf{k})$ of this phase space are the real
and imaginary parts of the Fourier amplitude of the incompressible
velocity field with an upper bound of the wave number $k$),
i.e., the dynamics of $\sigma_i(\textbf{k})$ satisfies
the Liouville theorem, by which an equipartition of energy, which was
considered as the conserved quantity, among $\sigma$s was then predicted
(c.f. Appendix \ref{APPsecPEDAGOGICALstat} for a pedagogical elaboration).
There are several reasons that the study
of the statistical mechanics of such idealized systems can be of
interest.\cite{Kraichnan-Montgomery80,Taylor97,Krommes02}  First is
that this can give
analytic (or semi-analytic)
predictions for the equilibrium statistics that can be used as a nonlinear
benchmark to test codes.  Such nonlinear analytic tests are rare and
thus valuable.  (This has been useful for fluid codes and, in plasma
physics, for particle-in-cell
codes,\cite{Langdon79,Birdsall85,KLO86,Krommes93,Krommes03_MC} and could be used for
continuum
kinetic codes as well.)  Second, such analytic spectra can also be
useful test cases for analytic theories of turbulence.
Equilibrium statistics has been shown to have subtle and deep relevance
to statistically nonequilibrium turbulence.
It has been used to  provide insights into two-dimensional (2D) guiding-center
plasma and 2-D vortex fluid
models,\cite{Taylor71,Edwards74,Kraichnan-Montgomery80}, and other plasma
models\cite{Montgomery80,Gang89}. More recently, it has provided
insights\cite{Taylor09} into the unexpected phenomena of spontaneous
``spin-up'' in bounded 2-D fluid turbulence
simulations.\cite{Clercx98,vanHeijst09}  (Interestingly, a current
research topic in the fusion field is spontaneous rotation observed in
tokamaks.\cite{rice:2008,solomon:2010})
The most well-known result from this approach may be the prediction of
inverse energy cascade in two dimensional turbulence by Kraichnan,
\cite{Kraichnan2D67} following which Frisch et
al.\cite{FrischMHD}~calculated the magnetohydrodynamic (MHD) absolute
equilibrium and
illustrated how the inverse cascade of magnetic helicity may help
explain the generation of large-scale magnetic fields in some
astrophysical systems. Another
example is how the concept of `partial thermalization' has recently been
used to understand some observed phenomena such as the `bottleneck' near
dissipation scales in Fourier space and the reduction of intermittency,
or its scaling, in physical space,\cite{FrischPRL08, ZhuTaylorCPL2010}
which emphasizes the persistence of some aspects of equilibrium
statistical mechanics in turbulence,
complementing the other side of our knowledge of the persistence of aspects of
cascade physics beyond the inertial range (see, e.g., Zhu \cite{jzzPRE05}).
Revisiting and further extending such powerful
tools to accumulate relevant knowledge and to examine the relevance to
definite realities is then important.  More recently, this
approach has been taken to analyze Hall MHD by Servidio et
al.,\cite{SMC08} finding that, among others, equipartition of kinetic
and magnetic energy predicted by Lee \cite{Lee1952} for Alfv\'enic MHD
turbulence no longer holds. Here we will take this paradigm to
investigate the gyrokinetic model of plasma turbulence.
The nontrivial new feature in our problem is that the integrations over the distributions are functional
integrals because of the extra dependence on velocity of the gyrokinetic variable.

More generally,
understanding the statistical mechanics of truncated gyrokinetics can
help shed light onto the general nature of nonlinear coupling in these
equations, and phenomena such as direct or inverse cascades.
A better understanding of nonlinear processes in gyrokinetics may also
help in the development of more effective sub-grid models for Large-Eddy
Simulations, and could improve understanding of the ultimate heating
mechanisms as the fluctuations cascade to very small spacial and velocity scales where collisional
dissipation occurs.\cite{Sch09Review,Schekochihin08}

Related to these sub-grid dissipation issues, the three-dimensional (3D) energy spectrum of
thermal fluctuations that we calculate here for a discretized Eulerian
gyrokinetic algorithm turns out to be closely related to the noise
spectrum calculated earlier for particle-in-cell (Lagrangian)
gyrokinetic algorithms.\cite{Nevins05}

In the two-spatial-and-one-velocity-dimension case, the negative temperature
state, leading to the condensation of the generalized energy at the
lowest modes,
indicates a generic feature of inverse energy cascade.
Comparisons are made with some
classical results, such as those of Charney-Hasegawa-Mima
in the cold-ion limit, though more generally the spectra
are modified by finite Larmor radius (FLR) effects which depend on the
temperature parameters. The shape of the statistical equilibrium for
gyrokinetics in three spatial and two velocity dimensions, where there
is just one conserved quantity, has a universal energy spectrum shape,
resulting from FLR effects.

In the main body we emphasize the general conceptual ideas with only
necessary details for illustration; specific mathematical calculations,
physical examples and other interesting digressions are referred to
the appendixes for further interests.

\section{Formulating the problem and calculating the absolute equilibria}
To be self-contained, here we very briefly introduce the nonlinear
gyrokinetic theoretical framework under which we will be working. We
won't review the complete history of the linear and nonlinear
gyrokinetic theories \cite{BH07review} but will just present the basic
ideas and results, borrowing from some of the treatment and notation of
Plunk et al.\cite{Gabe1,Gabe2}  There are several published derivations of
gyrokinetic equations with varied assumptions and techniques, including
recent papers with a tutorial emphasis.\cite{Howes06,Krommes10}
The starting point is the Boltzmann
equation for the particle distribution function
$f_s(\textbf{r},\textbf{v},t)$ for plasma species $s$ located at
$\textbf{r}$ moving with velocity $\textbf{v}$ at time $t$:
\begin{equation*}%\label{eq:Boltzmann}
\frac{\partial f_s}{\partial t} + {\bf v}\cdot \frac{\partial f_s}{\partial \textbf{r}} +
\frac{q_s}{m_s}({\bf E} + \frac{{\bf v}\times{\bf B}}{c}) \cdot \frac{\partial
  f_s}{\partial {\bf v}} = C[f_s]
\end{equation*}
Here the operator $C[f]$ accounts for the effects of collisions and the
particles with mass $m_s$ and charge $q_s$ are accelerated by the electric
($\textbf{E}$) and magnetic ($\textbf{B}$) fields, which are subject to the
classical Maxwell equations. The next step is introducing the
gyrokinetic ordering (which is fundamentally to focus on fluctuations
that are low frequency compared to the fast gyromotion of particles
around the magnetic field) and the resulting expansion parameter.
A key operation in the resulting equations is the average of any particular
quantity $\Psi$ around a ring of gyroradius $\rho$
perpendicular ($\perp$) to the magnetic field direction ($\parallel$)
% $\textbf{R}+\mbox{\boldmath$\rho$}$
surrounding the gyrocenter $\textbf{R}$:\cite{Note1}
\begin{equation}\label{eq:GA}
\langle \Psi \rangle_{\textbf{R}}=\frac{ \int \Psi(\textbf{r}) \delta( \textbf{r}_{\parallel} - \textbf{R}_{\parallel} ) \delta[ |\textbf{r}_{\perp}-\textbf{R}_{\perp}|-\rho(\textbf{R}) ] d^{3}\textbf{r} } {2\pi \rho(\textbf{R})}.
\end{equation}
Using a Fourier representation $\Psi(\textbf{r}) = \sum_k
\exp(-i\textbf{k} \cdot \textbf{r}) \hat{\Psi}_\textbf{k}$, and considering a
straight magnetic field for simplicity here, this becomes $\langle \Psi
\rangle_{\textbf{R}}=\sum_\textbf{k} \exp(-i \textbf{k} \cdot
\textbf{R}) J_0(k_\Perp \rho) \hat{\Psi}_\textbf{k}$, where $J_0$ is a Bessel
function.

Writing $\textbf{v}=\textbf{v}_{\perp}+v_{\parallel}\hat{\textbf{z}}$
and $f=F_{01}+h+h.o.t.$ (and suppressing the species subscript $s$ for
now), with $h.o.t.$ representing ``higher order terms,''
the resulting gyrokinetic equations for the case of slab geometry with
a homogeneous plasma in a straight equilibrium magnetic field
$\textbf{B}_0=B_0 \hat{\textbf{z}}$) is
\begin{equation*}%\label{eq:gyro}
\frac{\partial h}{\partial t}
+v_\parallel \hat{\textbf{z}} \cdot \frac{\partial h}{\partial \textbf{R}} +
\frac{c}{B_0} \left( \hat{\textbf{z}} \times
\frac{\partial \langle \chi \rangle_{\textbf{R}} }{\partial \textbf{R}}
\right) \cdot \frac{\partial h}{\partial \textbf{R}}
= {q} \frac{\partial \langle
\chi \rangle_{\textbf{R}}}{\partial t} \frac{F_{0}}{T_{0}},
\end{equation*}
complemented with the similar ordering-gyroaveraging treatment of the
Maxwell equations for the electrostatic potential $\varphi$ and the
perturbed vector potential $\textbf{A}$ which compose the gyrokinetic
potential $\chi=\varphi-\textbf{v}\cdot \textbf{A}/c$. Here the
collisional term is omitted. The zeroth and first order term $F_{01}$ is
in general taken to be the equilibrium Maxwell distribution ($F_0$)
multiplied by a Boltzmann factor, $\exp(-q \varphi / T_0) \approx 1 - q
\varphi / T_0$.
In what follows below, as in Plunk et al.,\cite{Gabe1,Gabe2} we will
work with the gyroaveraged, perturbed,
guiding center distribution function $g = h - F_0 q \langle \varphi
\rangle_R / T_0$, instead of with the
non-adiabatic component $h$, and for
simplicity we will focus on the case of electrostatic fluctuations
(neglecting magnetic fluctuations, $\textbf{A}=0$) with one particle
species governed by the gyrokinetic equation and the other species
having a Boltzmann response of some form (discussed below).

To make it easier to compare with other codes and theories that use a
variety of normalizations, and in particular to make it easier to take
the cold-ion limit in 2-D to compare with the Hasegawa-Mima equations,
we will use a generalized normalization for space and time scales based
on a reference temperature $T_r$, a reference sound speed $c_r =
\sqrt{T_r/m}$, and a reference gyroradius $\rho_r = c_r / \Omega_c$,
(here the mass $m$ and Larmor (cyclotron) frequency $\Omega_c = q B/mc $
are for the species that is governed by the gyrokinetic equation), but
still scale $v_\Par$ and the velocity dependence of $F_0$ and $g$ to $\vth =
\sqrt{T_0/m}$, where $T_0$ is the temperature of the gyrokinetic
species.  More specifically, we use the following normalizations and
definitions, with physical (dimensional) variables having subscript `p':
%\vspace{.5cm}
\begin{center}
\begin{tabular}{cccc}
$t = t_{\mbox{\scriptsize{p}}} c_r/L$ & $x = x_{\mbox{\scriptsize{p}}}/\rho_{r}$ & $y = y_{\mbox{\scriptsize{p}}}/\rho_{r}$ & $z = z_{\mbox{\scriptsize{p}}}/L$  \\
$v_{\Perp,\Par} = \frac{v_{\Perp,\Par,\mbox{\scriptsize{p}}}}{\vth}$ & $\phi = \phi_{\mbox{\scriptsize{p}}}\frac{q L}{T_r \rho_r}$ & $h = h_{\mbox{\scriptsize{p}}}\frac{\vth^3 L}{n_0 \rho_r}$  & $F_0 = \frac{F_{0\mbox{\scriptsize{p}}}\vth^3}{n_0}$
\end{tabular}
\end{center}
%\vspace{.5cm}
The equilibrium density and temperature of the gyrokinetic species of
interest are $n_0$ and $T_0$; the thermal velocity is $\vth =
\sqrt{T_0/m}$; $L$ is the reference macroscopic scale length
(\textit{i.e.}, system size), satisfying $\rho/L \ll 1$ for consistency
with gyrokinetic ordering.

In these normalized units, the Maxwellian background distribution
function is given by $F_0=\exp(-(v_\Perp^2+v_\Par^2)/2)/(2 \pi)^{3/2}$,
and the gyrokinetic equation for the gyroaveraged, perturbed, guiding
center density $g(\textbf{R}, v_\Par, v_\Perp,t)$ is given by
\begin{eqnarray}\label{eq:gyronorm}
\frac{\partial g}{\partial t}
+\rho_0 v_\parallel \frac{\partial g}{\partial z}
+ \left( \hat{\textbf{z}} \times \frac{\partial \langle \phi
  \rangle_{\textbf{R}} }{\partial \textbf{R}}
\right) \cdot \frac{\partial g}{\partial \textbf{R}}
= - \frac{v_\Par}{\rho_0}
\frac{\partial \langle \phi \rangle_{\textbf{R}}}{\partial z} F_{0},
     \nonumber \\
\end{eqnarray}
where $\rho_{0} = \rho_{th} / \rho_r = \vth/c_r = \sqrt{T_0/T_r}$ is the
thermal gyroradius $\rho_{th}$ of the gyrokinetic species normalized to
the reference gyroradius $\rho_r$.
% and the gyroaveraged potential in Fourier space is $\langle \phi
% \rangle_{\textbf{R}} = J_0(k_\Perp \rho_0 v_\Perp) \hat{\phi}(\textbf{k})$.
(Our normalization reduces to that
used in Plunk et~al.\cite{Gabe1,Gabe2}\ if we choose $T_r = T_0$ so
$\rho_0 = 1$, which in fact we will do in the 3-D case.)

The gyrokinetic equation expresses how the guiding centers evolve in
time due to parallel motion along the magnetic field, the gyro-averaged
$\textbf{E} \times \textbf{B}$ drift across the magnetic field (this is
the nonlinear term), and parallel electric field acceleration.  (Note
that the slow $\textbf{E} \times \textbf{B}$ drift of the guiding center
location $\textbf{R}$ is different than the rapid gyration velocity
$v_\Perp$ of a particle around its guiding center.)

This equation is closed by using the gyrokinetic quasi-neutrality
equation to determine the electrostatic potential, which in Fourier
space with these normalized units is given by
\begin{eqnarray}\label{eqQN3D}
\hat{\varphi}(\textbf{k},t) &=& \frac{\beta(\textbf{k})}{2 \pi}
   \int d^3v J_0(k_\Perp  \rho_0 v_\Perp)
   \hat{g}(\textbf{k},v_\Par,v_\Perp,t) \nonumber \\
  &=& \beta(\textbf{k}) \! \int_{-\infty}^\infty \! \! dv_\Par
   \int_0^\infty \! \! dv_\Perp v_\Perp
   J_0(k_\Perp  \rho_0 v_\Perp)
   \hat{g}(\textbf{k},v_\Par,v_\Perp,t), \nonumber \\
\end{eqnarray}
where
\begin{equation}\label{eqbeta}
\beta(\textbf{k})=\frac{2\pi}{\tau({\bf k})+\frac{T_r}{T_0}(1-
\hat{\Gamma}(k_\Perp^2 \rho_0^2))},
\end{equation}
$\hat{\Gamma}(k^2)=I_0(k^2)e^{-k^2}$ is an exponentially-scaled
modified Bessel function, $I_0(k^2) = J_0(i k^2)$,
and $\tau({\bf k})$ represents the shielding by the species that is treated as
having a Boltzmann response of some form, the choice of which depends on
physical situation.  If we are treating the ions gyrokinetically (such
as for ion-scale drift waves or Ion Temperature Gradient-driven
turbulence) and using an adiabatic approximation for electrons because
of their fast parallel motion relative to a typical frequency,
$k_\Par v_{te} \gg \omega$
(except for modes with $k_\Par = 0$), then $\tau(\textbf{k}) =
(T_r/T_e) (1 -
\delta_{k_\Par})$, where $T_e$ is the electron temperature and the
discrete Kronecker $\delta$ function ensures that the electrons do not
respond to zonal modes with $k_\Par = E_\Par = 0$.  If we are treating
electrons gyrokinetically (such as for small electron scale Electron
Temperature Gradient-driven turbulence) with an adiabatic approximation
for ions because $k_\perp v_{ti} \gg \omega$
(the $k_\perp = 0$ mode is
not driven by any nonlinearities in a periodic domain), then $\tau =
T_r/T_i$.

Finally, one can also consider a no-response model, $\tau=0$,
which in the 2-D cold-ion limit $T_0 \rightarrow 0$ leads to $\beta
\rightarrow 2 \pi / k_\Perp^2$, $J_0 \rightarrow 1$, and the
gyrokinetic equation reduces to 2-D hydrodynamics.

The above difference in zonal flow dynamics for ion vs. electron scale
fluctuations is responsible for a large enhancement in zonal flows for
ion-scale turbulence, so that zonal flows play a key role in the
saturation dynamics of ITG turbulence\cite{Hammett93,Dorland93,Cohen93}
and leads to the Dimits nonlinear shift in the critical
gradient.\cite{Dimits96,Dimits00}  It is also responsible for a
significant reduction in the effect of zonal flows for electron-scale
turbulence, so that they can get to larger amplitude than one would at
first expect from scaling from ion-scale
turbulence.\cite{Jenko2000,Dorland2000}

\subsection{\label{sec:2DGK} 2D Gyrokinetic absolute equilibria}
For a plasma in a two dimensional ($\partial/\partial z = 0$) cyclic
box, the collisionless gyrokinetic equation in wavenumber space reads
\begin{eqnarray}\label{eqg}
\partial_t \hat{g}(\textbf{k},v)=\hat{\textbf{z}}\times
\sum_{\textbf{p}+\textbf{q}=\textbf{k}} \textbf{p}J_0(p \rho_0
v)\hat{\varphi}(\textbf{p})\cdot \textbf{q} \, \hat{g}(\textbf{q},v)
\end{eqnarray} with the potential $\varphi$ determined by the
quasi-neutrality condition
\begin{eqnarray}\label{eqQN}
\hat{\varphi}(\textbf{k})
    =\beta(\textbf{k}) \int vdv J_0(k \rho_0 v) \hat{g}(\textbf{k},v),
\end{eqnarray}
where the subscript on $v_\Perp$ has been dropped and the parallel
velocity $v_\Par$ has been integrated out of the problem.

The only known rugged (still conserved after mode truncation) invariants
are the ``energy'' $\mathrm{E}= (1/2V) \int d^2\textbf{r} [(\tau + (T_r/T_0))
  \varphi^2- (T_r/T_0) \varphi \Gamma \varphi]$, and a parameterized set of
invariants related to the ``perturbed-entropy'' $\mathrm{G}(v)=(1/2V) \int
d^2\textbf{R} g^2$ (in these equations, $V$ is the volume (area) of
the integration domain and $\Gamma$ is a convolution operator in real
space given by the Fourier transform of $\hat{\Gamma}$).
(See Refs.(\onlinecite{Gabe1,Gabe2,Sch09Review}) and references therein for a
discussion of these conserved quantities and their interpretation.) In
Fourier space these become $\mathrm{E} =$ $\pi \sum_{\textbf{k}} |\phi_k|^2
/\beta({\textbf{k}})$ and $\mathrm{G}(v) =$ $ \sum_{\textbf{k}} |g({\bf
  k},v)|^2/2$.
As promised in the introductory discussion, following Lee,\cite{Lee1952}
in what follows we will
keep summations over only a finite subset $\mathbb{K}$ of all possible
wavenumbers $\textbf{k}$---the Fourier Galerkin
truncation.
\cite{Note2}
% End footnote
% ---and denote this by $\tilde{\sum}_{\bf k}$.
%
The Fourier modes in the lower half plane are determined by the reality
condition, $g(-\mathbf{k},v) = g^*(\mathbf{k}, v)$, so the state of a
system can be uniquely specified by the values of the real and imaginary
parts of the Fourier coefficients $g(\mathbf{k},v)$ for wavenumbers in
the upper half plane.  We will thus consider a further subset
$\mathbb{K}^+$, defined as the modes in $\mathbb{K}$ in the upper half
plane, which satisfy $k_y \ge 0$ if  $k_x>0$, or $k_y > 0$ if $k_x \le
0$
(see also Krommes and Rath\cite{Krommes03_MC}).  
All spectral sums will be expressed in terms of the finite set of
independent modes in $\mathbb{K}^+$, and we denote this summation by
$\tilde{\sum}_{\bf k}$.

We can
discretize Eq. (\ref{eqQN}) into
\begin{equation} \label{eqQN1D}
\tilde{\hat{\varphi}}(\textbf{k}) =  \beta(\textbf{k}) \sum_{i=1}^N w_i(k)
\hat{g}(\textbf{k},v_i),
\end{equation}
where $w_i(k) = m_i v_i J_0(k \rho_0 v_i)$, and $m_i$ is the weight of
velocity grid point $v_i$.
This discrete form can correspond to the case that $\hat{g}(\textbf{k}, v)$ is uniform on the lattice around node $i$; in general, it is used as a numerical approximation for the arbitrary distribution over $v$ as applied in the present continuum codes.\cite{Note3}
For a simple
midpoint integration rule on a grid that extends up to some maximum
velocity $v_{max} = v_N$, the weight is given by the grid spacing, $m_i
= \Delta v_i$.  (More general integration algorithms can also be
represented in this form.\cite{Note4})
With this velocity discretization, there are now $N+1$ conserved quantities,
given by the energy
$\tilde{E}= \sumkp 2 \pi \beta({\bf k})
\sum_{i,j}^{N}w_i \hat{g}^*(\textbf{k},v_i) w_j
\hat{g}(\textbf{k},v_j)$ and the en\-tropy-related quantities
$\tilde{G}_i= \sumkp |\hat{g}(\textbf{k},v_i)|^2$.

So, with the common belief of the applicability
of Gibbsian statistical mechanics or Jaynes' \cite{Jaynes57I,
  Jaynes57II} idea of ``statistical mechanics as a form of statistical
inference", we have the distribution function $\sim \exp
\{-\tilde{\mathcal{S}}\}$, where $\tilde{\mathcal{S}}$ is a linear
combination of conserved quantities, which will be written as
\begin{eqnarray}\label{eqH-main}
\tilde{\mathcal{S}} = && \sum_{i=1}^N \alpha_i \tilde{G}_i+\alpha_0\tilde{E}.
\end{eqnarray}
Here, the $\alpha_i$ are the ``(inverse) temperature parameters''
introduced as Lagrangian multipliers to form the constant of the motion.
The Gibbs measure can be shown to be conserved by the flow as a generalized Liouville theorem, the incompressibility of the flow of the phase points in the
hyperplane spanned by the real and imaginary parts of the Fourier
modes.\cite{Lee1952}
Note that, importantly, conservation laws and the Liouville theorem are inherited, which, with some more assumptions (such as ergodicity) makes the discrete system possible to produce a Gibbs ensemble.
A pedagogical illustration on
the Gibbs canonical distribution for this system can be found in Appendix
\ref{APPsecPEDAGOGICALstat}. 

As this is a multivariate Gaussian
distribution, one can numerically invert the matrix of the quadratic
form
$\mathcal{S} = \sum_{i,j}[\delta_{ij} \alpha_i + \alpha_0 2 \pi \beta(\textbf{k})w_i(k) w_j(k)]\hat{g}(\textbf{k},v_i)\hat{g}^{\ast}(\textbf{k},v_j)$,
 but, actually, using the Sherman-Morrison formula (see Appendix
\ref{APPsecShermanMorrison}), one can write down the $N\times N$
covariances
\begin{eqnarray}\label{eq:covariance}
c_{i,j}(\textbf{k}) &=&
    \langle g^*(\textbf{k},v_i) g(\textbf{k},v_j) \rangle /2\nonumber \\
  &=&\frac{\delta_{i,j}}{2 \alpha_i}-
  \frac{\alpha_0 \pi \beta(\textbf{k}) w_i
    \alpha_i^{-1}w_j\alpha_j^{-1}}{1+\alpha_02\pi \beta(\textbf{k}) \sum_l w_l^2
    \alpha_l^{-1}}.
\end{eqnarray}
The spectral energy density $D(\textbf{k})$ then can be calculated as
follows:
\begin{eqnarray}\label{eq:elecspec}
D(\textbf{k})&&= \frac{\pi}{\beta(\textbf{k})}
     \langle \left| \hat{\varphi}(\textbf{k}) \right|^2 \rangle
   = 2 \pi \beta(\textbf{k}) \sum_{i,j} w_i w_j  c_{ij}\nonumber\\
&&= \frac{\pi \beta(\textbf{k})\sum_l w_l^2
    (\alpha_l)^{-1}}{1+\alpha_0 2\pi \beta(\textbf{k}) \sum_l w_l^2
    \alpha_l^{-1}}.
\end{eqnarray}
The isotropic energy spectrum then is $E(k) \sim k D(\textbf{k})$.  The
spectral density of the perturbed entropy $\mathrm{G}_i$ is
\begin{eqnarray}\label{Gspec}
\mathrm{G}_i(\textbf{k}) &=& \frac{1}{2}
   \langle \left| \hat{g}(\textbf{k},v_i) \right|^2 \rangle =
   c_{i,i}(\textbf{k}) \nonumber \\
   &=& \frac{1}{2 \alpha_i} \left[ 1 - \frac{\alpha_0 2\pi
  \beta(\textbf{k}) w_i^2 \alpha_i^{-1}}{1+\alpha_0 2\pi \beta(\textbf{k})
  \sum_l w_l^2 \alpha_l^{-1}} \right],
\end{eqnarray}
which expresses the nearly equipartition on Fourier modes
when the second term in the brackets has negligible contribution.\cite{Zhu20103}
(Note that $D({\bf k})$ and $G_i({\bf
k})$ depend on wavenumber not only through $\beta({\bf k})$ but also
through $w_i=w_i(k)$.)

The temperature parameters are determined by the values of the
invariants $\tilde{G}_i$ and $\tilde{E}$.
Note that these give the
invariants as nonlinear functions of the temperature
parameters $\alpha_0$ and $\alpha_i$, so this requires numerical
solution or further analytic approximations to invert.

\subsubsection*{Discussion concerning 2+1D gyrokinetic spectra}
\label{Discussion-2D}
To gain insight into the behavior of these spectra, we will explore them
in various simplifying limits, such as the small and large-$k$ limits
and the cold-ion Hasegawa-Mima limit.  We will also plot example spectra
for particular values of the $\alpha_0$ and $\alpha_i$ parameters.

The gyroaveraging by plasma particles through Eq. (\ref{eq:GA})
(finite Larmor radius effects) introduces several special functions
whose asymptotic behaviors help shape the energy spectrum.
For convenience, let's write here the asymptotic recipes for these special
functions:
\begin{center}
\begin{tabular}{ccc}
$k \to 0,$ &  $\! J_0(k \rho_0 v_i) \approx 1 - k^2 \rho_0^2 v_i^2 / 4,$ &
  $\, \, \hat{\Gamma}(k^2 \rho_0^2) \approx 1-k^2 \rho_0^2$  \\
$k\to \infty,$ & $J_0^2(k \rho_0 v_i) \sim \frac{2 \cos^2(k v_i -
    \pi/2)} {(\pi k v_i)},$ &
   $\hat{\Gamma}(k^2 \rho_0^2) \sim \frac{1}{\sqrt{2\pi}k \rho_0}$
\end{tabular}
\end{center}

The simplest limit to consider first is $\rho_0 = \sqrt{T_0/T_r}
\rightarrow 0$, i.e., the cold-ion limit considered by Hasegawa and
Mima in their study of drift-wave turbulence \cite{HMpof78} (see
Appendix~\ref{APPsecGK2FLUID} for more details).  The Hasegawa-Mima
equation coincides with the Charney equation for geophysical
flows\cite{Charney71} formulated earlier, so it is also called the
Charney-Hasegawa-Mima (CHM) equation.

In this limit, $J_0(k \rho_0 v)
\rightarrow 1$, and $w_i(k) = m_i v_i J_0(k \rho_0 v)
\rightarrow m_i v_i$ so the factor of $\sum_l w_l^2(k) / \alpha_l$ in
Eq.~(\ref{eq:elecspec}) becomes independent of $k$ and can just be taken
as a constant, which we will define as $1/(2 \pi^2 \bar{\alpha})$.
In the expression for $\beta({\bf k})$ in
Eq.~(\ref{eqbeta}), we use $\hat{\Gamma}(k^2 \rho_0^2) \rightarrow 1 -
k^2 \rho_0^2$ and find $\beta({\bf k}) \rightarrow 2 \pi / (\tau +
k^2)$.  For comparison with Hasegawa and Mima, we set the reference
temperature to the electron temperature, $T_r = T_e$, (the
electrons are the adiabatic species for ion-scale drift waves)
and neglect the $\delta_{k\Par}$ factor and associated zonal flow effects
\cite{Note5}
in the expression for $\tau$ so that $\tau=1$.  Note
that our normalized $k = k_\Perp = k_{\Perp,p} \rho_s$, where
$k_{\Perp,p}$ is the physical perpendicular wavenumber, and $\rho_s =
\sqrt{T_e/m_i}/\Omega_{ci}$ is the ion sound radius.  (Alternatively,
one can consider these as equations for electron-scale turbulence where
the role of ions and electrons is reversed: the ions are adiabatic and a
cold electron limit is used, in which case the normalizing length is an
``electron sound radius'', $\rho_{se} = \sqrt{T_i/m_e}/\Omega_{ce}$.)

The result is that the isotropic energy spectrum given by
Eq.~(\ref{eq:elecspec}) reduces in the cold ion limit to
\begin{equation}
E(k) \propto k D(k) \propto \frac{k}{\bar{\alpha} (1 + k^2) + 2 \alpha_0},
\label{eq:2dspec}
\end{equation}
where $\bar{\alpha}$ and $\alpha_0$ are coefficients that are determined
by the values of the invariants $E$ and $G_i$.  Note that this is of the
same form of a 2-parameter family of spectra as in Charney-Hasegawa-Mima
(CHM) or 2-D Euler absolute equilibrium, $E(k) \propto k / (\alpha_{CHM}
+ \beta_{CHM} k^2)$ (Appendix \ref{APPsecGK2FLUID}).

\begin{figure}
\includegraphics[width=\columnwidth]{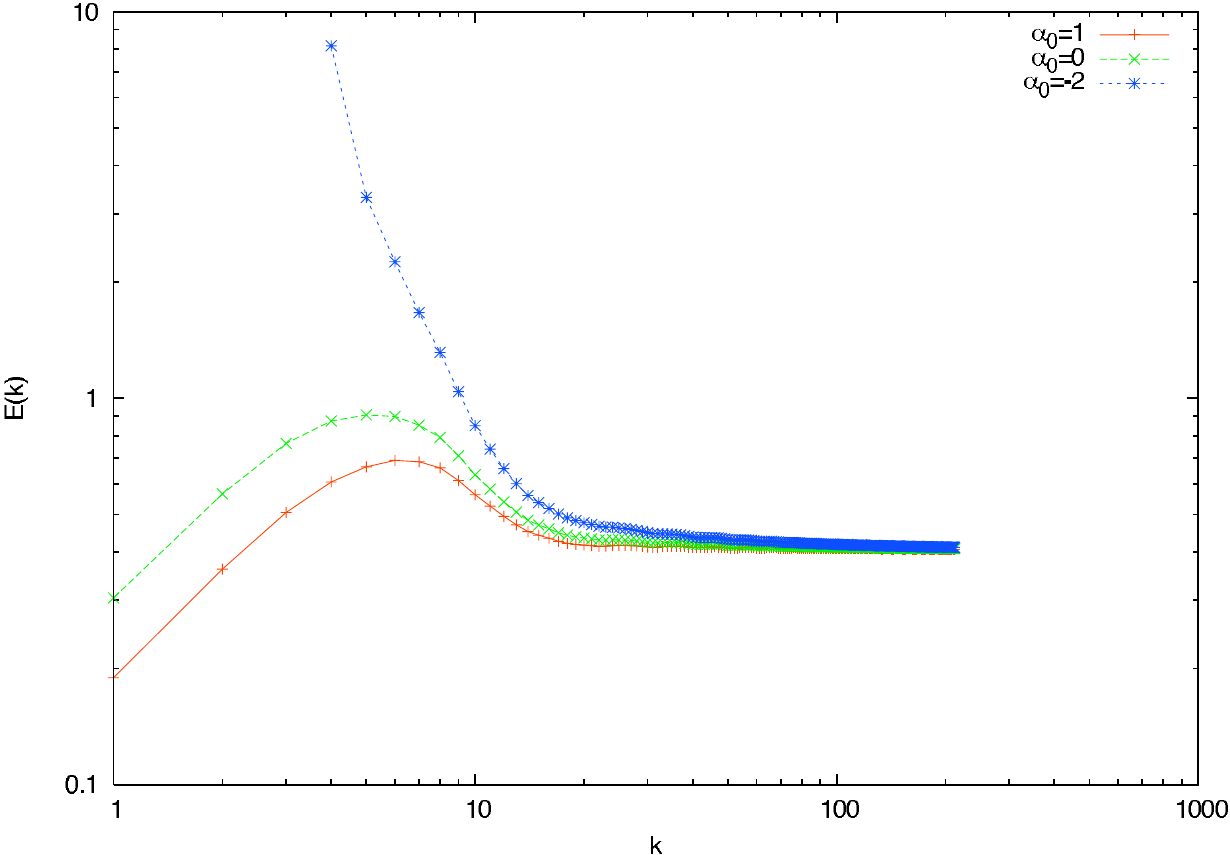}% Here is how to import EPS art
\caption{\label{fig:Ekaa2} (Color online) Example spectra for various values of
  $\alpha_0$, with $\alpha_i=10^3\exp\{v_i^2/2\}$ for $i > 0$, and $\rho_0 =10^{-1}$, $\tau=1$. Negative $\alpha_0$ state can occur that correspond to condensation
  of most of the energy into the longest wavelength modes.}
\end{figure}

A remarkable feature of this type of spectrum is that if $\alpha_{CHM}$
or $\beta_{CHM}$ are negative, corresponding to a negative temperature,
then the denominator has the opportunity of tending to zero leading to
the energy condensation  at the lowest or highest wave
numbers.
Energy condensation to the lowest modes (one example is shown in
Fig.~\ref{fig:Ekaa2}) would
indicate an inverse cascade of energy following the argument of
Kraichnan for the 2D Euler equation \cite{Kraichnan2D67} (see also
Hasegawa and Mima \cite{HMpof78}, and, Fyfe and
Montgomery.\cite{FMpof79})  It then seems that the inverse cascade of
energy
in 2D gyrokinetics could be
quite a generic feature: Recent relevant theoretical arguments and numerical simulation results \cite{Sch09Review, Gabe2,TomoPRL09} are consistent with this.

Instead of the cold-ion limit, we now consider the more general case of
warm ions (for ion-scale turbulence, or warm electrons for
electron-scale turbulence), and for simplicity let us take
$T_e=T_i=T_r$ (so that $\rho_0=1$) and neglect zonal flow effects (so
that $\tau=1$).
In the limit $k \gg 1$, we have $\beta(\textbf{k}) \rightarrow \pi$, and the
magnitude of $w_i(k)^2 \propto J_0^2(k v_i)$ will be bounded by $C/(k
v_i)$ for some constant $C$.
Assuming positive $\alpha_i$ for $i > 0$,
we find that the denominator in Eq.~(\ref{eq:elecspec}) approaches 1 for
large $k$, so $D(k) \sim 1/k$ and $E(k) \sim k D(k) \sim k^0$.  This
could give a larger tail for gyrokinetics than for CHM, which
has $E_{CHM}(k) \sim 1/k$ (for $\beta_{CHM}>0$).  
In this same $k \gg 1$ limit,
Eq.~(\ref{Gspec}) simplifies to $G_i({\bf k}) = 1/(2 \alpha_i)$, which
corresponds to equipartition of the generalized entropy. We plot the spectra over the reachable wave number regime with given temperatures just to sketch the physical picture without even bothering to accurately calculate the realizable wave number bounds but only with estimations sufficient to directly illustrate the problem. (Another way to think about the problem and do the corresponding plots is to take the bounds of wavenumbers to be prescribed and then realizable temperature parameters are determined accordingly. Detailed computations and illustrations of more example spectra with possible physical discussions are given in Appendix \ref{APP:example-inv-spectra} for those who are interested.) Actually, much is already known from the knowledge of absolute equilibria of 2D Euler \cite{Kraichnan2D67} and Hasegawa-Mima \cite{HMpof78}, though it may still be helpful to give some general physical picture, especially the finite Larmor radius effects, with some example spectra as shown in Fig.\ref{fig:Ekaa2}: The $\alpha_i$ for $i>0$ were set by
$\alpha_i^{-1} = 10^{-3} \exp(-v_i^2/2)$. We take $N=40$ with $v_i$ homogeneously collocated between $0$ and $V=3$, and, $\mathbb{K}=\{\textbf{k}|1\leqslant k_{x,y}\leqslant 150\}$ for positive temperatures cases while $\mathbb{K}=\{\textbf{k}|4\leqslant k_{x,y}\leqslant 150\}$ for a negative temperature case. Some details, including the values of $m_i$ ($=1$ here), are not important, and reasonable changing of them (as a re-normalization of the variables) won't affect our results. Here, as $E(k) \sim kD(\textbf{k})$, the low-$k$ equipartition range have $E(k) \sim k$ and $E(k) \sim k^{0}$ for large $k$, both of which can be easily checked with the asymptotic recipe of the special functions given in the beginning of this subsection. The transition in between represents the FLR effects, the details of which also depend on the details of the temperature parameters. Other temperature parameter values may change the $k$ ranges where such asymptotic behaviors can be realized; or, if the truncation wave numbers in the computations $k_{min}$ and $k_{max}$ (which may be relevant to some characteristic, such as the collisional, scales in real physical systems) were not chosen properly, we would not be able to reach such asymptotic behaviors. 
% End of section describing Fig. 1

A negative value of $\alpha_0$ corresponds not only to an enhancement of
energy at larger spatial scales (low $k$) but also to an enhancement of
fluctuations at larger velocity scales, as given by
Eq.~(\ref{eq:covariance}).  If $\alpha_0=0$, then this equation indicates
that different velocity grid points $i \not= j$ are uncorrelated, while
making $\alpha_0$ more negative will increase the correlation length in
the velocity, particularly at low $k$.  If collisions are included, they
will cause dissipation at both small velocity scales and small spatial
scales (through FLR effects corresponding to classical
diffusion),\cite{Abel08} so the inverse cascade found here in 2D will
tend to reduce both forms of dissipation.

We end up this sub-section by remarking that, since the CHM limit absolute equilibrium statistics has already been verified by numerical experiment,\cite{FMpof79} our theoretical calculation for gyrokinetic is then also, to some degree, endorsed.

\subsection{\label{sec:3DGK} 3+2D Gyrokinetic absolute equilibria}
The mathematical treatment for the calculation of Galerkin truncation absolute equilibrium is basically the same for 3+2D and 2+1D cases. However, some brief remarks about the system and the conserved quantity, followed by some technical details in the calculation are still necessary. 

In the full gyrokinetic equation with three spatial and two velocity
dimensions, notice that two linear terms, from parallel motion along the magnetic field and from the parallel electric field acceleration, are simply added to the equation for the $(2+1)$-D case, without changing anything about the nonlinear term. While the gyroaveraged E cross B drift conserves $E$ and $G$ separately, the parallel motion makes them talk to each other and combines them into another conserved quantity, the
generalized energy $W = E + W_{g0}$ (see
Refs.(\onlinecite{Gabe2,Sch09Review,Schekochihin08,Krommes94}) and references
therein for further discussions of this quantity):
\begin{equation} \label{W3d}
W =
\int \frac{d^3\mathbf{r}}{2 V} [(1+\tau) \varphi^2-\varphi \Gamma \varphi]
+ \int d^3 v  \int \frac{d^3\mathbf{R}}{2 V} \frac{g^2}{F_0}.
\end{equation}
(For simplicity in all of our 3+2D work, we will set the reference
temperature $T_r$ used in normalizations to the temperature of the
kinetic species $T_0$, so $\rho_0=1$.)  The Fourier Galerkin truncated
form of the generalized energy is
\begin{equation*}
S = 2 \sumkp \Bigg\{ \frac{\pi}{\beta(\textbf{k})} |\hat{
  \varphi}(\mathbf{k})|^2 + \pi \int \!\!\!\!\! \int
v_{\perp}dv_{\perp}dv_{\parallel}
 \frac{|\hat{g}(\mathbf{k},{\bf v})|^2}{F_0}
\Bigg\}.
\end{equation*}
We will discretize velocity space in a way that makes it easy to
reduce the previous 2D spatial + 1D velocity results to the 3D
spatial + 2D velocity results here.  Specifically, we will discretize
velocity integrals as
\begin{eqnarray}
\int d^3 v g({\bf v}) &=& 2 \pi \int_0^\infty d v_\Perp
\int_{-\infty}^{\infty} d v_\Par v_\Perp g(v_\Par,v_\Perp) \nonumber \\
&\approx& 2 \pi \sum_i^N m_i v_{\Perp,i} g({\bf v}_i),
\end{eqnarray}
where $i$ now indexes over all points ${\bf v}_i = (v_{\Perp i},v_{\Par
i})$ in the 2D velocity space grid.  For a logically rectangular mesh
in $(v_\Perp,v_\Par)$ there would be a total of $N=N_{v_\Perp}
N_{v_\Par}$ grid points, with $N_{v_\Perp}$ points in the perpendicular
velocity direction and $N_{v_\Par}$ in the parallel velocity direction.
(As in the 1D velocity case, for a simple midpoint integration
algorithm, $m_i$ is the weight of the $i$'th velocity cell, $m_i =
\Delta v_{\Perp,i} \Delta v_{\Par,i}$, while more generally the weights
$m_i$ and grid point locations $(v_{\Perp,i},v_{\Par,i})$ can be chosen
to give high-order Gaussian quadrature.)  The 2D velocity generalization
of Eq.~(\ref{eqQN1D}), the discretized quasineutrality equation to
determine the potential, now reads $\hat{\phi}(\textbf{k}) =
\beta(\textbf{k}) \sum_i w_i(k_\Perp) \hat{g}(\textbf{k}, \textbf{v}_i)$,
where $w_i(k_\Perp) = m_i v_{\Perp,i} J_0(k_\Perp v_{\Perp,i})$.
% , and $k =
% |\textbf{k}| = k_\Perp$ in the gyrokinetic expansion in
% $\epsilon=\omega/\Omega_c \rightarrow 0$ since $k_\Par \sim \epsilon
% k_\Perp$.

We then can calculate the absolute equilibria following the same
procedure as in the 2D case, but now only one inverse temperature
parameter $\gamma$ shows up in the canonical distribution $\sim
\exp\{-\gamma S\}$.  Using the above velocity discretization for
$\tilde{\cal S} = \gamma S = \gamma (E+W_{g0})$ gives
\begin{eqnarray*}
\tilde{\mathcal{S}} &=& \gamma \sumkp 2 \pi \beta(\textbf{k})
   \Sigma_{i,j}w_i
   \hat{g}^*(\textbf{k},{\bf v}_i) w_j
   \hat{g}(\textbf{k},{\bf v}_j)   \nonumber \\
&& + \gamma \Sigma_{i=1}^N 2 \pi m_i v_{\Perp,i} \sumkp
|\hat{g}(\textbf{k},{\bf v}_i)|^2/F_0(v_i)
\end{eqnarray*}

We note from this that a negative temperature is not realizable any
more.  Comparing this expression for the 3D $\tilde{\cal S}$ with the 2D
result in Eq.~(\ref{H2D}), we see that they become identical if we make
the substitutions $\alpha_0 = \gamma$ and $\alpha_i = 2 \pi \gamma m_i
v_{\Perp,i} / F_0(\textbf{v}_i)$.  All of the 2-D results thus
generalize to the 3-D case with these variable substitutions.  For
example, the electrostatic component of the spectral energy density
in Eq.~(\ref{eq:elecspec}) becomes
\begin{equation}\label{Dspec3d}
D(\mathbf{k}) = \frac{1}{2 \gamma}
\left[  \frac{\beta(\textbf{k}) \sum_i m_i v_{\Perp,i} F_0(\vec{v}_i)
      J_0^2(k_\Perp v_{\Perp,i})}{1
   + \beta(\textbf{k}) \sum_i m_i v_{\Perp,i} F_0(\vec{v}_i) J_0^2(k_\Perp
   v_{\Perp,i})} \right]
\end{equation}

In the small lattice size limit \cite{Zhu20103}, where we can use $2 \pi \sum_i m_i v_{\Perp,i}
F_0(\vec{v}_i) J_0^2(k_\Perp v_{\Perp,i}) \approx \int d^3 v
F_0(\vec{v}_i) J_0^2(k_\Perp v_{\Perp,i}) = \Gamma_0(k_\Perp^2)$,
the electrostatic potential spectral density becomes
\begin{equation}\label{phi3d}
\langle \left| \phi_{\textbf{k}} \right|^2 \rangle =
\frac{\beta(\textbf{k})}{\pi} D(\mathbf{k}) = \frac{1}{\gamma}
\frac{\Gamma_0(k_\Perp^2)}{(\tau + 1 - \Gamma_0(k_\Perp^2)) (\tau + 1)}
\end{equation}

The shape of this spectrum
is consistent
 with the discrete-particle thermal
noise spectrum for gyrokinetic PIC codes calculated by one of us
previously, as given in Eq.~(5) of Ref.(\onlinecite{Nevins05}), which
reduces to the above result in the limit where numerical details such as
spatial filtering and finite differencing
\cite{Note7}
are ignored by setting $S_G(\textbf{k})=1$ and $d_\Par(\textbf{k})= 1$,
and by taking the $\tau=1$ limit in our expression.  (The thermal
spectrum in Ref.(\onlinecite{Nevins05}) was calculated for the case of
one gyrokinetic species and one adiabatic species, as also assumed in
the present paper, and also accounted for various numerical factors as used
in typical PIC codes.  The first calculations of the discrete-particle
thermal noise spectrum for gyrokinetic particle codes are in
Refs.(\onlinecite{Krommes93,Hu-Krommes94}) and were for the case where
all species were treated gyrokinetically.)

Ref.(\onlinecite{Nevins05}) found good agreement between this analytic
thermal spectrum and the fluctuation spectrum in a PIC code in a
noise-dominated regime, providing support for the calculation done here.
Readers interested in a discussion of noise in numerical schemes are referred to Appendix \ref{APPsecNoise}.

\subsubsection*{Relevance to 3D plasma turbulence}

There are several interesting features of the 3D spectrum in
Eq.~(\ref{phi3d}).  Note that it is independent of $k_\Par$, i.e., the
equilibrium spectrum corresponds to equipartition in $k_\Par$, so
presumably the nonlinear dynamics of a turbulent system should tend to
drive cascades to high $k_\Par$.  (Gyrokinetics assumes $k_\Par \ll
k_\Perp$, so there is a limit to how far this spectrum can extend within
this model.)  Also note that even with the finite-Larmor radius
averaging in gyrokinetics, the electrostatic potential spectrum falls
relatively slowly at high $k_\Perp$ since $\Gamma_0 \sim C/k_\Perp$, so
the electrostatic energy spectrum is flat at high wave number,
$E_{\phi}(k_\perp) \propto k_\Perp |\phi(\textbf{k})|^2 \sim k_\Perp^0$.

For ion-scale non-zonal flows with adiabatic electrons, the
long-wavelength limit of Eq.~(\ref{phi3d}) is
$\langle | \phi_{\bf k} |^2 \rangle = 1 / (2 \gamma)$
(setting $\tau = T_r/T_e = T_i/T_e = 1$
for simplicity).  But for zonal flows, which have $k_y=k_z=k_\Par=0$ and
thus have $\tau=0$ (see the discussion after Eq.~(\ref{eqbeta})), the long
wavelength limit is  $\langle | \phi_{ZF,{\bf k}} |^2 \rangle = 1 / (\gamma
k_x^2)$ (here the ``ZF'' subscript refers to the zonal flow component of
the potential), so the amplitude of long-wavelength zonal flows is enhanced
relative to other nearby modes by a factor of $\sim 1/k_x$.
(While the resulting zonal potential blows up as $k_x \rightarrow 0$,
the shearing rate $\propto d v_y/dx \propto d^2 \phi_{ZF}/dx^2 \propto k_x^2
\phi_{ZF,\textbf{k}} \propto k_x$ remains well-behaved.)
However, one of us, GWH, tends to believe that this
enhancement in the 3-D statistical
equilibrium is interesting but by
itself is probably not enough to explain the observed importance of
zonal flows in ITG turbulence, since there are very few zonal modes
compared to the many other modes with $k_y \neq 0$ or $k_\Par \neq 0$.
The importance of ITG zonal flows is probably due to other effects, such
as the way in which the lack of adiabatic electron response causes an
enhancement of the secondary instabilities\cite{Rogers00} (or
related parametric instabilities) that drive zonal flows.

However, much stronger enhancement of zonal flows can exist in the 2-D
absolute equilibrium of Eq.~(\ref{eq:elecspec}) where negative
$\alpha_0$ can strongly enhance modes with $\tau=0$.
The mechanism for this enhancement is related in a way to
the enhancement of zonal flows in secondary/parametric instabilties.
This 2-D equilibrium effect might be related to the enhancement of zonal
flows in an actual turbulent plasma, if there are regions of the
turbulent spectrum where the parallel dynamics is slow compared to the
nonlinear decorrelation rate $k_\Par v_t \ll \Delta \omega_{NL} \sim
k_\Perp v_{E \times B}$ and so act in a quasi-2D manner.  However, there
will also be competition from 3D effects, which limits the inverse
cascade and tends to push the spectrum towards equipartition in
$k_\Par$.

The 2D and 3D gyrokinetic absolute equilibrium results may also provide
insight into other aspects of driven non-equilibrium gyrokinetic
turbulence, such as the directions of turbulent cascades in
$(k_\Par,k_\Perp)$. The inverse cascade found in 2D may imply that in
regions of a turbulent spectrum where $k_\Par v_t \ll k_\Perp v_{E\times
  B}$, then the interactions may be quasi-2D and undergo an inverse
cascade to smaller $k_\Perp$, simultaneously with a cascade to higher
$k_\Par$ (towards equipartition in $k_\Par$), until the parallel
dynamics becomes competitive with nonlinear terms, $k_\Par v_t \sim
k_\Perp v_{E\times B}$.  At this point it might then switch to a forward
cascade to higher wavenumber, but along a path in $(k_\Par,k_\Perp)$
space such that $k_\Par v_t \sim k_\Perp v_{E \times B}$.  Thus this
supports the critical balance hypothesis suggested for gyrokinetic
turbulence in Refs.(\onlinecite{Sch09Review,Schekochihin08}), that the
turbulence will primarily cascade along a path in wave number space that
has parallel linear time scales comparable to perpendicular nonlinear
time scales, similar to critical balance ideas in astrophysical Alfv\'en
turbulence in Refs.(\onlinecite{Goldreich95,Goldreich97}).  Further
analysis of gyrokinetic statistical equilibria may lead to more specific
insights.

There are other more subtle physics, such as the bottleneck
and its associated weakening of intermittency growth
issues,\cite{jzzCPL06} as proposed to be explained as partial
thermalization by Frisch et al. \cite{FrischPRL08}
For example, as the Fourier transform is
linear, the physical-space field of the Fourier Galerkin truncated
absolute equilibria would also be Gaussian, whose residual may result in
a resistance in the departure from Gaussian (intermittency) for the
turbulence fluctuations.\cite{ZhuTaylorCPL2010} Before examining the details of collision and
wave-particle interaction mechanisms, so far we unfortunately are not
able to say anything more on this for the plasma turbulence.
Nevertheless, such considerations emphasize the importance of implementing the correct collision operators (which is necessary in many physical situations) and in
  interpreting the numerical data.

\section{Conclusion and further remarks}
Here we have extended previous work on statistical equilibria of 2D and
3D hydrodynamics and MHD to the case of higher-dimensional gyrokinetics.
Previous work in hydrodynamics found that there was a profound
difference between 2D and 3D, because the existence of 2 invariants in
2D lead to the existence of negative temperature equilibrium states with
most of the energy condensing into the longest wavelengths in the system
(related to the inverse energy cascade in 2D turbulence), while in 3D
there was only a single invariant resulting in energy equipartition
among Fourier modes (related to the forward cascade of energy to small
scales in 3D turbulence).

For gyrokinetics in the limit of 2 spatial and 1 velocity dimension (2+1D),
we have worked out the Gibbs equilibrium in the presence of $N+1$
invariants (where $N$ is the number of velocity grid points) and find
that, like 2D hydrodynamics, this can also exhibit negative temperature
states where much of the energy condenses to the longest wavelengths in
the system.  For a range of typical parameters explored so far, 2+1D
gyrokinetics exhibits a very strong inverse cascade.
At high $k_\Perp$, the 2D gyrokinetic energy spectrum has
an asymptotically-flat tail, $E(k_\Perp) \sim k_\Perp^0$, which is
enhanced relative to the high $k_\Perp$ limit of Hasegawa-Mima's thermal
spectrum, $E(k_\Perp) \sim 1/k_\Perp$.  The amplitude of this tail in
gyrokinetics is found to depend sensitively on the ratio of $G_i$ to
energy.

We also calculated the statistical absolute spectrum for
Fourier-truncated gyrokinetics in the full 3 spatial and 2 velocity
dimensions, and found that the result was equivalent to earlier thermal
noise spectra calculated for particle-in-cell gyrokinetics, indicating
that the random phase and amplitude of shielded Fourier components of
the distribution function in a continuum representation is related to
the random position and weights of shielded particles in the
Klimontovich representation of a PIC code.  The resulting 3-D
gyrokinetic spectrum corresponds to equipartition in $k_\Par$, and even
with all of the finite-Larmor radius averaging in gyrokinetics, the
electrostatic potential spectrum only falls relatively slowly at high
$k_\Perp$, so $E_{\phi}(k_\perp) \propto k_\Perp |\phi(\textbf{k})|^2
\sim k_\Perp^0$.

As described in the introduction, statistical equilibria spectra as
calculated here have several useful purposes.  In particular, they
provide an analytic nonlinear test for benchmarking of gyrokinetic
codes, which could be pursued in future work.  They may also provide
insights into certain aspects of the nonlinear dynamics in driven,
non-equilibrium gyrokinetic turbulence simulations.
For example, in regions of the turbulent spectrum where the parallel
linear dynamics is slow compared to the nonlinear decorrelation rate,
$k_\Par v_t \ll k_\Perp v_{E\times B}$, then the interactions may behave
in a quasi-2D behavior, which can cause an inverse cascade to smaller
$k_\Perp$ in general, and in particular can strongly enhance the ITG
zonal flows because of the lack of adiabatic electron shielding for ITG
zonal flows.  But these may be offset by the tendency towards
equipartition of the spectrum in $k_\Par$, so that eventually $k_\Par
v_t \sim k_\Perp v_{E\times B}$ and parallel linear dynamics becomes
competitive with nonlinear perpendicular dynamics.  In this region of
wavenumbers, the turbulent cascade would then switch to a forward
cascade to higher $|\textbf{k}|$, along a path where parallel and
perpendicular dynamics remain comparable and so stay full 3D, consistent
with the critical balance hypothesis for gyrokinetic turbulence
suggested in Refs.({\onlinecite{Sch09Review,Schekochihin08}}).
There are various directions in which the present work could be extended
in the future that may further help in understanding plasma behavior in
actual experiments, such as extensions to include a kinetic treatment of
all particle species, electromagnetic fluctuations, and the effects of
magnetic curvature and grad-$B$ drifts in toroidal geometry.

% \vspace{-1em}
\begin{acknowledgments}
% \vspace{-1em}
We acknowledge
many
colleagues for helpful interactions related to this
work, especially G. Plunk, W. Dorland, J. Krommes,
W. M. Nevins and T. Tatsuno.
We also thank
W. Dorland, M. Zarnstorff, and S. Prager for their support.  This work
was supported by the U.S. Department of Energy through the Center for
Multiscale Plasma Dynamics at the University of Maryland, Contract
No. DE-FC02-04ER5478, the SciDAC Center for the Study of Plasma
Microturbulence, and the Princeton Plasma Physics Laboratory by DOE
Contract No. DE-AC02-09CH11466.

\end{acknowledgments}

\appendix

\section{Pedagogical illustration of the calculation of the canonical ensemble} \label{APPsecPEDAGOGICALstat}
The line of reasoning presented by T.-D. Lee\cite{Lee1952} regarding how
to apply a statistical mechanics approach to hydrodynamics and MHD can
be straightforwardly extended to the higher dimensional gyrokinetic case
considered here.  We briefly summarize that line of reasoning here,
which also serves to explain the notation that we use.
Consider a system governed by Eqs.~(\ref{eqg}) and (\ref{eqQN1D}) (with
Eq.~(\ref{eqg}) evaluated at the same velocity grid points as used in
Eq.(\ref{eqQN1D})).
The state of a
system at a particular time can be specified by a vector ${\bf g}$
in an extended phase space of dimension $N_k N$, where $N_k$ is the
number of Fourier modes and $N$ is the number of velocity grid points.
The elements of ${\bf g}$ are $\hat{g}(\textbf{k},v_i)$, the complex
amplitude of Fourier modes $\textbf{k} \in \mathbb{K}^+$ at velocities
$v_i$, where $\mathbb{K}^+$ is the set of independent modes $\textbf{k}$
in the truncation $\mathbb{K}$ that are in the upper half plane.
% , $(k_x>0, k_y
% \ge 0)$ or $(k_x \le 0, k_y > 0)$.  (The Fourier modes in the lower half
% plane are determined by the reality condition, $g(-\mathbf{k},v_i) =
% g^*(\mathbf{k}, v_i)$.)
One can
consider an ensemble of many such systems, and define the function
${\cal P}({\bf g},t)$ that gives the probability of a system being in
state ${\bf g}$ at time $t$.  For continuous dynamics, this satisfies a
conservation law $\partial_t {\cal P} + \partial_{\bf g} \cdot \left(
\dot{\bf g} {\cal P} \right) = 0$ where an over-dot is used to denote a
time derivative so $\dot{{\bf g}}$ is given by Eq.~(\ref{eqg}).  A
generalized Liouville theorem holds for these equations, i.e., the flow
in this extended phase space is incompressible, $\partial_{{\bf g}}
\cdot \dot{{\bf g}} = 0$, because for a given value of $\textbf{k}$, the
right hand side of Eq.~(\ref{eqg}) vanishes if $\textbf{p}=\pm\textbf{k}$
(because $\textbf{q} = \textbf{k}-\textbf{p}$ means $\textbf{p} \times
\textbf{q}$ vanishes),
and thus
also vanishes if $\textbf{q} = \pm \textbf{k}$.
(In other words, the rate
of change $\dot{g}(\textbf{k})$ at any instant in time depends only on
the amplitude of other modes $\hat{g}(\textbf{p})$ with $\textbf{p} \neq
\textbf{k}$.)

Since a Liouville theorem holds, standard results and assumptions
from statistical mechanics can be applied to these equations. A
generalized Liouville equation holds, $\partial_t {\cal P} + \dot{{\bf
    g}} \cdot \partial_{{\bf g}} {\cal P} = 0$, i.e., the probability
${\cal P}({\bf g}(t),t)$ is constant on a moving trajectory in this
extended phase-space.  Looking for a time-independent statistical steady
state, we take an equal probability for all points along a trajectory's
path.  Assuming that the dynamics are sufficiently mixing and an ergodic
hypothesis holds, so that a trajectory samples all possible points on a
hyper-surface in phase-space constrained only by the invariants, leads
to the micro-canonical ensemble given by ${\cal P} = C \delta(E-E_0)
\Pi_{i=1}^N \delta(G_i-G_{i0})$, where $E=E({\bf g})$ and $G_i({\bf g})$
are the previously given expressions for the energy and entropy
invariants, which are functions of ${\bf g}$, $E_0$ and $G_{i0}$ are the
values of those invariants (set by initial conditions), and
$\Pi_{i=1}^N$ indicates repeated multiplication over all possible
velocity points $i$.

As is well known, for systems with a large number of degrees of freedom,
many features of a micro-canonical ensemble are often well-approximated
by a Gibbs canonical ensemble, ${\cal P} = Z^{-1} \exp(-{\cal S})$ where
${\cal S}$ is a linear combination of conserved quantities, which in
this case is ${\cal S} = \alpha_0 E + \sum_i \alpha_i G_i$, $\alpha_0$
and the $N$ values of $\alpha_i$ are the ``(inverse) temperature
parameters'', and $Z$ is a normalization coefficient such that $\int
d{\bf g} {\cal P}({\bf g}) = 1$.  One way\cite{Jaynes57I,Jaynes57II}
to derive this is to choose ${\cal P}$ to maximize the Liouville
phase-space entropy $S_L = - \int d{\bf g} {\cal P}({\bf g}) \log({\cal
P}({\bf g}))$ (i.e., choose ${\cal P}$ to be as uniformly
distributed as possible) subject only to constraints on the average
values of the invariants (this leads to the Lagrange multipliers
$\alpha_0$ and $\alpha_i$ in the canonical ensemble).  For example,
the constraint on the ensemble-averaged value of the energy is $E_0 =
\langle E \rangle = \int d{\bf g} {\cal P}({\bf g}) E({\bf g})$.

[Note that the fact that a Liouville theorem is satisfied is an
important part of justifying the maximum-entropy approach of the
previous paragraph, as it means that the probability distribution ${\cal P}$
can be constant along trajectories in these coordinates, which is not
necessarily true in other coordinates.  For example, something that is
uniformly distributed in $x$ is not uniform in $x^3$.]

Inserting the expressions for the energy and entropy invariants into the
the expression for $\cal S$ gives
\begin{eqnarray}\label{H2D}
{\cal S} &=& \alpha_0 \sumkp
% {\sum}_{{\bf k} \in \mathbb{K}^+}^{}
2 \pi \beta(\textbf{k}) \sum_{i,j} \hat{g}^*(\textbf{k},v_i) w_i(k) w_j(k)
\hat{g}(\textbf{k},v_j)  \nonumber \\
&& + \sum_{i=1}^N \alpha_i
\sumkp
% {\sum}_{{\bf k} \in \mathbb{K}^+}
|\hat{g}(\textbf{k},v_i)|^2
\end{eqnarray}

With a little rearrangement, the Gibbs canonical distribution becomes
\begin{eqnarray}\label{eqH}
{\cal P}
&\propto& \exp\left(-\frac{1}{2}
\sumkp
%{\sum}_{{\bf k} \in \mathbb{K}^+}
    \textbf{g}^*(\textbf{k}) \cdot
    \textbf{M}(\textbf{k}) \cdot \textbf{g}(\textbf{k}) \right),
\end{eqnarray}
This is of the form of a multivariate Gaussian distribution, where the
elements of the $\textbf{k}$-dependent, $N \times N$ matrix $\textbf{M}$
are given by $M_{ij} = \delta_{ij} 2 \alpha_i + \alpha_0 4 \pi \beta(\textbf{k})
w_i(k) w_j(k)$, and $\textbf{g}(\textbf{k})$ is the $N$-dimensional
vector of the velocity-indexed values of the complex amplitudes
$\hat{g}(\textbf{k},v_i)$.

Expressing ${\bf g}$ in terms of its real and imaginary parts, ${\bf
  g}({\bf k}) = {\bf g}_R({\bf k}) + i {\bf g}_I({\bf k})$, note that
the sum over wavenumbers in Eq.~(\ref{eqH}) can be
written as $\sumkp
{\bf g}_R({\bf k}) \cdot {\bf M}({\bf k}) \cdot {\bf g}_R({\bf k})
+
\sumkp
{\bf g}_I({\bf k}) \cdot {\bf M}({\bf k}) \cdot {\bf g}_I({\bf k})$
since ${\bf M}$ is real, so
the real and imaginary parts of ${\bf g}$ are uncorrelated and
have the same co-variance,
$\langle g_R({\bf k},v_i) g_R({\bf k}, v_j) \rangle =
\langle g_I({\bf k},v_i) g_I({\bf k}, v_j) \rangle = c_{ij}({\bf
  k})$,
where $c_{ij}$ are the elements of the co-variance matrix $C({\bf
k})$ given by the inverse of $\bf M$, i.e., ${\bf C}(\textbf{k}) = {\bf
  M}^{-1}(\textbf{k})$.

\section{Calculating the covariance matrix using Sherman-Morrison formula} \label{APPsecShermanMorrison}
In principle, once that ${\bf M}({\bf k})$ in Eq.~(\ref{eqH}) is known,
one can calculate the co-variance matrix
% $ \langle
% {\bf g}^*({\bf k}) {\bf g}({\bf k}) \rangle \equiv {\bf C}({\bf k})$ from
% the inverse of ${\bf M}$, i.e.
${\bf C}(\textbf{k}) = {\bf M}^{-1}(\textbf{k})$, and one can then
calculate various statistical
properties of interest, such as the energy spectrum of fluctuations.
However, ${\bf M}$ is in general a dense matrix, so at first it looks
like this may require a numerical treatment to invert it. To make
analytic progress, one can initially consider the limit $\alpha_0=0$, in
which case ${\bf M}$ is diagonal and easily invertible.  One can then do
a matrix series expansion for small $\alpha_0$ and discover that it is
possible to sum the result to all orders in $\alpha_0$ because of the
special tensor product form of the coefficient of $\alpha_0$ in ${\bf
M}$. This turns out to be a special case of the general
Sherman-Morrison formula.

In linear algebra, suppose $A$ is an invertible square matrix and $u$,
$v$ are vectors and that $1 + v^T A^{-1}u \neq 0$, then the Sherman-Morrison
formula reads \cite{BartlettAMS51}
$$(A+uv^T)^{-1} = A^{-1} -
\frac{A^{-1}uv^T A^{-1}}{1 + v^T A^{-1}u}.
$$
To derive the covariance matrix ${\bf C} = {\bf M}^{-1}$ in
Eq.~(\ref{eq:covariance}), we use the definition of ${\bf M}$ given after
Eq.~(\ref{eqH}).
% , $M_{i,j} =\delta_{i,j} \alpha_i + \alpha_0 2 \pi \beta w_i
% w_j$.
Thus in the Sherman-Morrison formula, the elements of $A$ are
$a_{ij}=2 \alpha_i \delta_{ij}$, and we can set
$u_i = \alpha_0 4 \pi \beta w_i$ and $v_i = w_i$.
Then the elements of $A^{-1}uv^TA^{-1}$ are
$x_{ij}=\alpha_{0}2\pi \beta(\textbf{k}) w_i \alpha_i^{-1}w_j\alpha_j^{-1}$,
and $v^T A^{-1} u=\alpha_{0}2\pi \beta(\textbf{k}) \sum_i \sum_j w_i w_j
\alpha_i^{-1} \delta_{ij}=\alpha_{0}2\pi \beta(\textbf{k}) \sum_i w_i^2
\alpha_i^{-1} $. So, we have Eq.~(\ref{eq:covariance}).

\section{From gyrokinetics to fluids: recovering the Charney-Hasegawa-Mima equations}\label{APPsecGK2FLUID}
We first briefly reproduce the derivation of the Charney-Hasegawa-Mima
(CHM) equations from gyrokinetics by Plunk et al.\cite{Gabe2}\ with slight variation: From Eqs. (\ref{eqg}) and (\ref{eqQN}) we have
\begin{eqnarray}\label{eqphi}
\partial_t \hat{\varphi}(\textbf{k})=\beta(\textbf{k}) \sum_{\textbf{p}+\textbf{q}=\textbf{k}} \hat{\textbf{z}}\times \textbf{p}\cdot \textbf{q}  \hat{\varphi}(\textbf{p}) \nonumber \\
\times \int vdv J_0(k \rho_0 v) J_0(p \rho_0 v) \hat{g}(\textbf{q},v).
\end{eqnarray}
In the cold ion limit $\rho_0 \rightarrow 0$ (as described in Sec.~
\ref{Discussion-2D}), the first kind zeroth order Bessel functions
reduce to unity, and $\beta(\textbf{k})$ to $2 \pi / (\tau+k^2)$,
and then, with substitution of the quasi-neutrality Eq.~(\ref{eqQN}) in
the second line of (\ref{eqphi}), gyrokinetics reduces to CHM.
In physical space, it reads
\begin{equation}\label{eq:chm}
\partial_t(\tau - \nabla^2)\phi=\hat{{\bf z}}\times \bnabla \phi \cdot\bnabla (\nabla^2\phi)
\end{equation}
This is the inviscid (the collision operator in Plunk et
al.\cite{Gabe2}\ also vanishes after integration over velocity by
particle conservation) CHM equation. The scale to which
gradients were normalized in these equations
corresponds to the Rossby deformation radius in quasi-geostrophic
turbulence, or to the sound Larmor radius, $\rho_s$, in a plasma.  We
have left a $\tau$ dependence in these equations for generality, as the
two-dimensional Euler equation can be obtained in the case $\tau = 0$ (the
no-response model).

There are two invariants of the CHM/Euler equation that are relevant to
our discussion, referred to as energy and enstrophy (although their
physical interpretation depends on the specific scale of interest):
\begin{subequations}
\label{eq:chm-invar-defs}
\begin{align}
& E_{CHM} = \frac{1}{2}\int \frac{d^2{\bf r}}{V}[\tau \phi^2 + |\bnabla\phi|^2] \\
& Z_{CHM} = \frac{1}{2}\int \frac{d^2{\bf r}}{V}[\tau |\bnabla\phi|^2 + (\nabla^2\phi)^2].
\end{align}
\end{subequations}
This leads to absolute equilibrium energy spectra of the form $E(k)
\propto k D(k) \propto k/(\alpha_{CHM} + \beta_{CHM} k^2)$, where
$\alpha_{CHM}$ and $\beta_{CHM}$ will be determined by the values of
these two invariants, via $E_{CHM} = 2 \tilde{\sum}_{\bf k} D(k)$ and
$Z_{CHM} = 2 \tilde{\sum}_{\bf k} k^2 D(k)$.
The first invariant, $E_{CHM}$, is formally the reduced energy, $E$, of
gyrokinetics in the cold-ion limit. The enstrophy, $Z_{CHM}$, however is
new and deserves further inspection of its origin.

The CHM equation can be written as $\partial n/\partial t = \hat{\bf z}
\times \nabla \phi \cdot \nabla n$, where the potential is determined
from the guiding-center density $n = 2 \pi \int dv v g$ by inverting
$(\tau - \nabla^2) \phi = n$. The nonlinear term on the RHS of CHM has
the property that it can be multiplied by either the density $n$ or the
potential $\phi$ and then will vanish when integrated over all
space. This leads to the two standard energy and enstrophy invariants
used for the CHM equations.  However, in gyrokinetics where we keep a
finite, non-zero temperature, the velocity and wavenumber dependence in
the Bessel functions in the second line of Eq.~(\ref{eqphi}) introduces
extra non-local position and velocity scale interactions that mean that
$Z_{CHM}$ is no longer conserved.  Due to these FLR effects,
gyrokinetics instead has a set of invariants that hold at each velocity,
$G(v) \propto \int d^2 R g^2({\bf R},v)$, while CHM had an additional
invariant proportional to $\int d^2 R (\int dv v g)^2$.  This new CHM
invariant is not representable as a combination of the gyrokinetic
$G(v)$ and $E$ invariants.  One way to think of this is to note that CHM
depends only on the velocity integral of $g$ through $n = 2 \pi \int dv
v g$, so the CHM dynamics are independent of any details of the velocity
structure of $g$, thus allowing an additional invariant that is not
present in gyrokinetics because of its FLR effects.  

\section{\label{APP:example-inv-spectra} Example 2D Spectra for Specified
Initial Conditions\cite{Notejzz}}

Here we consider a numerical {\it gedankenexperiment}, in which a
gyrokinetic code is operated in the 2+1D limit and is initialized with
perturbations concentrated near some initial wavenumber $k_0$ but with
no other forcing.  Those perturbations will then interact nonlinearly
and scatter energy to other wave numbers, while preserving certain
invariants of the motion.  Presumably the spectrum will eventually reach
a statistical steady state, and here we make plots of the energy spectra
expected from canonical equilibria corresponding to some sample initial
conditions.  This helps provide further insight into the nature of these
equilibria.

Before making these plots, we first consider some of the properties of
spectra and the relationship between the invariants and the $\alpha_i$
parameters in more detail.
For positive $\alpha_0$ and $\alpha_i$ ($i > 0$), then the factor in
brackets in Eq.~(\ref{Gspec}) is close to unity for all wavenumbers, and
one can sum over all wavenumbers to find $G_i = N_{\bf k}/(2 \alpha_i)$
(where $N_{\bf k} \approx \pi (k_{max}/k_{min})^2$ is the number of Fourier
modes), which can be used to determine $\alpha_i$ in terms of the
conserved $G_i$.  Eq.(\ref{eq:elecspec}) can be summed over all
wavenumbers to determine the total energy and then determine
$\alpha_0$. For fixed positive values of $\alpha_i$, the energy is a
monotonically decreasing function of $\alpha_0$, so if the energy is
sufficiently large (for given values of the $G_i$), then $\alpha_0$ must
go negative to produce a ``negative temperature'' state.  If we
perturbatively use $\alpha_i \propto 1/G_i$ to evaluate the energy
spectrum, and assume a Maxwellian velocity distribution for the
fluctuations so $G_i = \int d^2 R g^2({\bf R},v)/2 \propto \exp(-v^2)$,
then the commonly occurring factor $\sum_l w_l^2 \alpha_l^{-1} \propto
\sum_l v_l^2 J_0^2(k v_l) \exp(-v^2)$ is a monotonically decreasing
function of $k$, as is $\beta(\textbf{k})$, so if $\alpha_0$ goes negative in the
denominator of Eq.~(\ref{eq:elecspec}), it will preferentially enhance
the energy in the low-$k$ part of the spectrum.
(If the denominator of Eq.~(\ref{eq:elecspec}) gets sufficiently close
to zero for some wavenumbers, then the factor in square brackets in
Eq.~(\ref{Gspec}) could differ from unity and alter the relationship
between $G_i$ and $\alpha_i$ assumed here.)
Note that the realizability constraint that the energy spectrum be
non-negative in this case means that the limiting value of $\alpha_0$
for this set
of $\alpha_i$'s is $\alpha_{lim} = - [2 \pi \beta({\bf k}) \sum_l
w_l^2(k) \alpha_l^{-1}]^{-1}$ evaluated at $k=k_{min}$.  (If the enhancement
of ion-scale zonal flows due to the lack of electron response is accounted for,
then this would strongly increase the value of $\beta$ for the
zonal modes, reduce the magnitude of the limiting value of $\alpha_0$,
and strongly enhance the amplitude of zonal flows.) 

Returning to the numerical {\it gedankenexperiment}, consider an initial
perturbation of the form
\begin{equation}
g(\vec R, v, t=0) = \cos(k_0 R_y) \frac{e^{-v^2/2}}{2 \pi}
J_0(k_0 v)
\label{model-ic}
\end{equation}
(here we set $\rho_0=1$ for simplicity), as a model that has some
characteristics of the drive by drift-wave types of instabilities.  This
initial condition models what happens if a linear source term $-v_{E \times
  B} \cdot \nabla F_0$ (representing instabilities that drive drift-wave
gyrokinetic turbulence) had been turned on in the gyrokinetic equation
for a time of order $L/c_r$ in the presence of a background density
gradient $\nabla F_0 = - \hat{x} F_0 / L$, where the eddy has a bi-normal
wavenumber $k_y = k_0$.  (In an actual code, a small amount of energy
must initially be put in other Fourier modes as well, because a single
Fourier mode does not interact with itself nonlinearly.) The energy and
entropy invariants corresponding to this initial condition are $E =
\beta(k_0) \hat{\Gamma}_0^2(k_0^2) / (8 \pi)$ and $G_i = \exp(-v_i^2)
J_0^2(k_0 v_i) / (16 \pi^2)$.  

Given the specified values of the energy and entropy invariants, it is
not analytically easy in general to invert the equations to determine
the corresponding temperature parameters $\alpha_0$ and $\alpha_i$,
because the energy and entropy are nonlinear functions of the
temperature parameters, as discussed after Eq.~(\ref{Gspec}).  
That is, the entropy invariants are related to the covariance matrix by
$G_i= 2 \tilde{\sum}_{\bf k} c_{i,i}(\textbf{k})$, and the energy
is related by $E = 2 \pi \tilde{\sum}_{\bf k}
|\phi_k|^2/\beta({\bf k}) = \tilde{\sum}_{\bf k} 4 \pi\beta(\textbf{k})
\sum_{i,j}^{N}w_i w_j c_{i,j}(\textbf{k})$, where the wavenumber sums
are over the independent set $\mathbb{K}^+$ and $c_{i,j}({\bf k})$ is a
nonlinear function of the temperature parameters as given by
Eq.~(\ref{eq:covariance}).

A
code was written to numerically carry out the inversion using a
nonlinear root solver based on Powell's method and Broyden's quasi-Newton
algorithm in the {\it minpack} software package.\cite{minpack}  To aid
in finding a root, a variable transformation was used for $\alpha_0$ to
ensure that during the search $\alpha_0$ never exceeded the lower limit
set by realizability constraints that the energy spectra be non-negative.
The results in this section used a uniformly spaced 2-D wavenumber
grid, \cite{Note6} 
where the set of retained Fourier modes is $\mathbb{K} = \{ {\bf k} | \,
k_{min} \le |{\bf k}| < k_{max} + \Delta k/2 \}$, with $k_{min} = \Delta
k = 0.1$, $k_{max}=10$, and the number of modes is $N_{\bf k} = 31,576$.
A uniformly-spaced velocity grid was used with $N=40$ points, equally
spaced from $\Delta v/2$ to $v_{max}=3+\Delta v/2$, with $m_i = \Delta v
= 3 / N$.  The background plasma temperatures were set to $T_0=T_e=T_r$ so
$\rho_0 = 1$ and an adiabiatic species response factor of $\tau=1$ for
simplicity, neglecting possible enhancements of zonal flows. 

\begin{figure}
\includegraphics[width=\columnwidth]{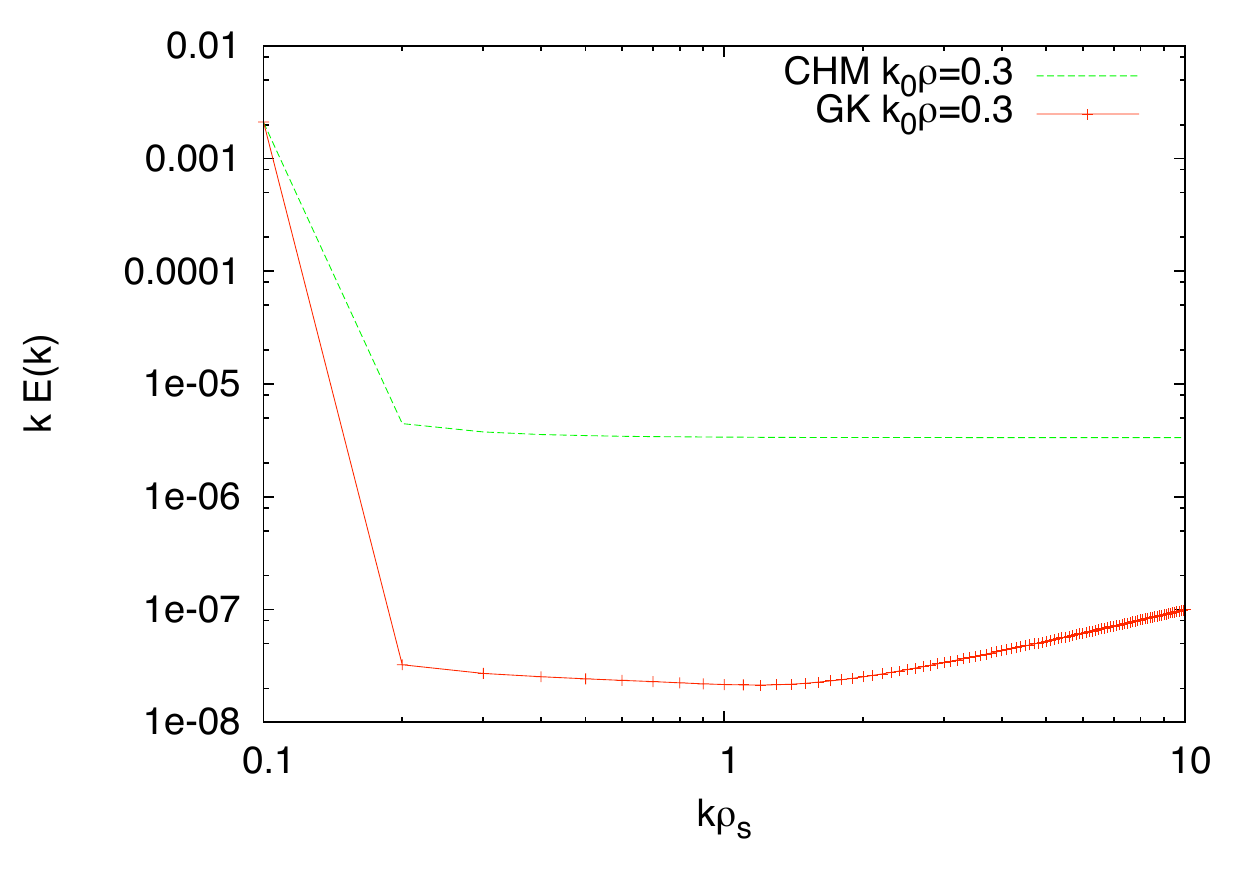}
\caption{\label{gkstat-gk-chm} (Color online) Spectra for 2+1D gyrokinetics (GK) and
  for the Charney-Hasegawa-Mima (CHM) equations, corresponding to the
  model initial conditions with energy initially at $k_0 \rho_s = 0.3$.
  Both spectra show a
  significant transfer of energy to larger scales, resulting in a negative
  temperature state with most of the energy condensed into the longest
  wavelength in the domain.}
\end{figure}

Fig.~(\ref{gkstat-gk-chm}) shows the gyrokinetic equilibrium spectrum
that results from these initial conditions with $k_0 = 0.3$.  (The
wavenumbers in the figures refers to the physical $k_p$, where the
normalized $k = k_p \rho_r$, and the reference gyroradius is set to
$\rho_s = \sqrt{T_e/m_i}/\Omega_{ci}$ for comparison with the cold-ion
Hasegawa-Mima
drift-wave equations.)  This figure also shows the spectrum given by the
Charney-Hasegawa-Mima equations for these same initial conditions.
\cite{Note8}
Both the
gyrokinetic and CHM spectra show strong transfer of energy
to large scales relative to the initial location of the energy at $k_0 = 0.3$, though
there is more of a tail in CHM case.  Both the gyrokinetic and CHM
equations result in a negative temperature state ($\alpha_0 < 0 $ for
the gyrokinetic case\cite{Note9})
with most of the energy condensed into the longest
wavelengths in the domain.

\begin{figure}[htb]
\includegraphics[width=\columnwidth]{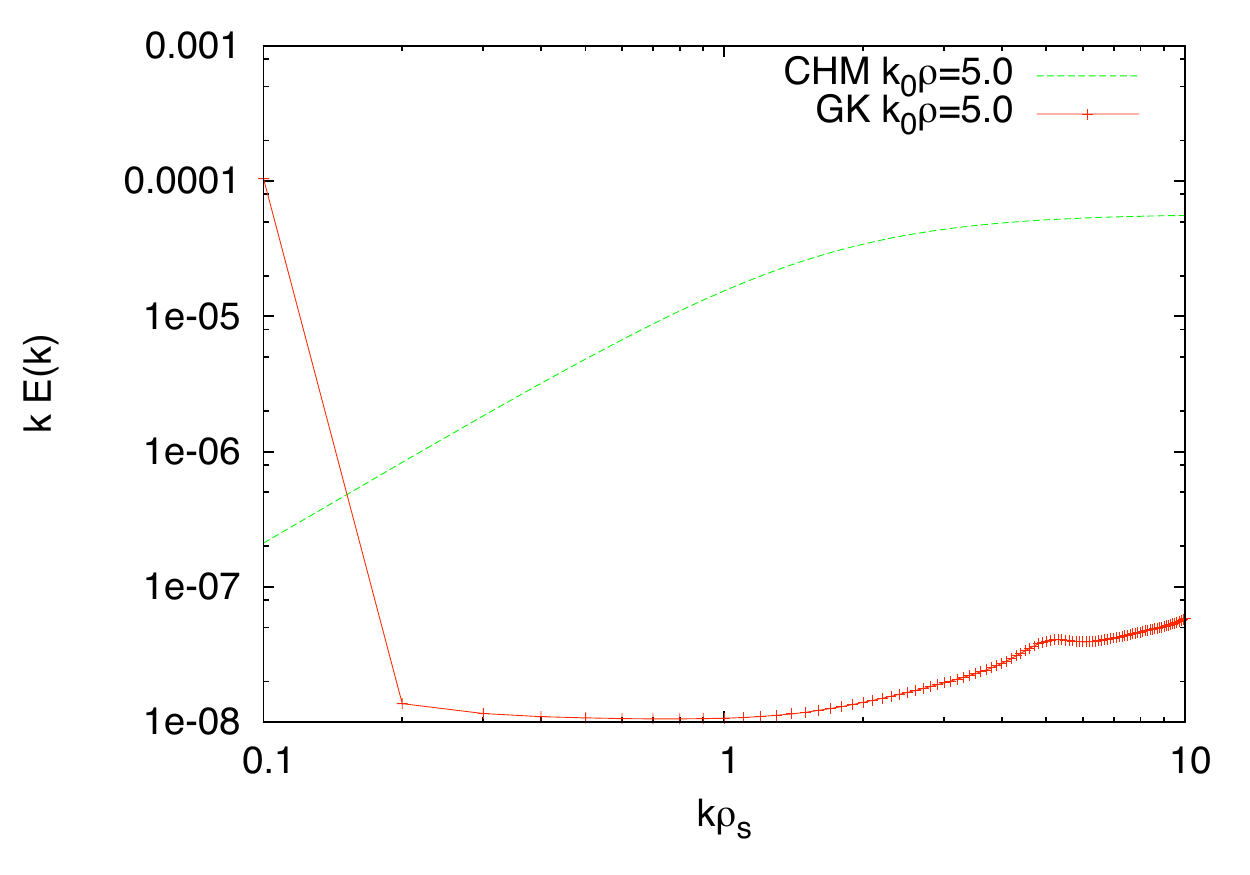}
\caption{\label{gkstat-gk-chm-5} (Color online) Spectra for 2+1D gyrokinetics
and Charney-Hasegawa-Mima, like Fig.~(\ref{gkstat-gk-chm}) except the
energy is initially at a higher wavenumber of $k_0 \rho_s = 5.0$.}
\end{figure}

Fig.~(\ref{gkstat-gk-chm-5}) is similar to Fig.~(\ref{gkstat-gk-chm})
except that the energy is initially at a higher wavenumber of $k_0 =
5.0$.  There are now significant differences between the gyrokinetic and
Hasegawa-Mima spectra, with gyrokinetics still showing a very strong
transfer to large scales while the energy remains primarily at
higher $k$ in the
Hasegawa-Mima case.  This is because the cold-ion Hasegawa-Mima
equations have an additional invariant (see Appendix
\ref{APPsecGK2FLUID}), the enstrophy (the mean squared vorticity), which
is not conserved by the general warm-ion gyrokinetic equations because
of FLR effects in the Bessel functions.  
(This is related to the fact that although the 2+1D gyrokinetic spectrum
in Eq.~(\ref{eq:2dspec}) for the $T_0/T_r \ll 1$ regime has the same
2-parameter form as the Charney-Hasegawa-Mima (CHM) spectrum, the
relationship between those 2 parameters and the invariants is different
for gyrokinetics than for CHM, because CHM has an additional invariant at
$T_0=0$ that doesn't exist in gyrokinetics with non-zero $T_0$.)
The initial wavenumber of $k_0 =5$ in Fig.~(\ref{gkstat-gk-chm-5}) 
is sufficiently close to the truncation wavenumber $k_{max}=10$ that
there is not much room for the enstrophy density $\propto k^2 E(k)$ to
transfer to higher wavenumber, thus inhibiting how much transfer
of energy to larger scales can occur in the Hasegawa-Mima equations.

\begin{figure}[htb]
\includegraphics[width=\columnwidth]{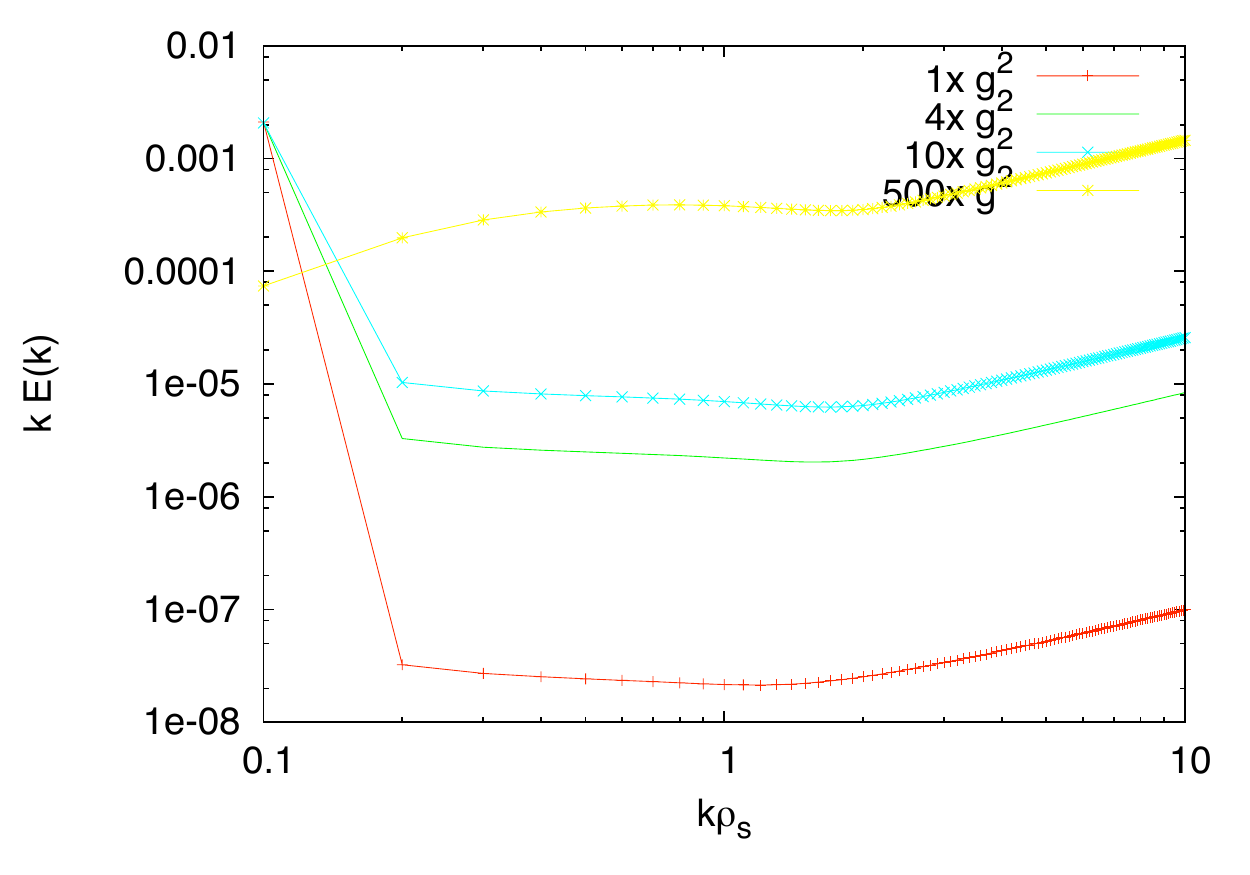}
\caption{\label{gscan-g} (Color online) Spectra for 2+1D gyrokinetics for
initial conditions with the energy at $k_0 \rho_s = 0.3$ like
Fig.~(\ref{gkstat-gk-chm}), but with the value of the entropy invariants
$G_i$ enhanced relative to the model initial conditions by factors of
$1 \times$, $4 \times$, $10 \times$, and $500 \times$.}
\end{figure}

It is possible to increase the size of the tail in the 2+1D gyrokinetic
spectrum by increasing the amplitude of the $G_i$ relative to the
energy, as shown in Fig.~(\ref{gscan-g}).  Considering a long-wavelength
initial condition ignoring FLR effects, this can occur if a component is
added to $g({\bf R},v,t=0)$ that oscillates in velocity so that it makes
no contribution to the potential $\propto \int dv v g$, but does enhance
$G_i = \int d^2 R \, |g({\bf R},v_i)|^2$.  This can model the effects of
temperature gradients in the background $F_0$ that drives the initial
perturbation, or the build up of large values of $G_i$ in a long
turbulence simulation without adequate dissipation because of the
entropy balance relationships,\cite{Krommes94,Nevins05} thus leading to
bottleneck problems.\cite{FrischPRL08}  Note that the
dependence of the tail on the enhancement of $G_i$ is a strongly
nonlinear function.
\cite{Note10}
In the limit of very large $G_i/E$, the spectrum
will approach equipartition among Fourier modes, $E \propto k$.  One
can also consider how the spectrum depends on the assumed velocity grid
spacing $\Delta v$. From numerical results, confirmed by analytic
scalings, one finds that as $\Delta v$ goes to zero, with fixed values
of $E$ and $G_i$, that $\alpha_0 \propto - 1 / \Delta v$ goes to
negative infinity (while $\alpha_i \rightarrow$ constant for $i>0$),
corresponding to a negative temperature state with all of the energy in
the lowest $k$ mode, so $E(k) = 0$ for all $k > k_{min}$.

\section{Thermal noise spectra in numerical schemes}\label{APPsecNoise}
In the 3+2D results in Sec.~(\ref{sec:3DGK}), we worked out the
electrostatic spectrum, Eq.~(\ref{phi3d}) and showed that the shape agrees
with earlier results for the spectrum in a PIC code.  Here we show
that the magnitude agrees as well, with the proper relation between
certain quantities in a continuum code and a PIC code.
Begin by defining a weighted mean-square average of the distribution
function $\overline{g^2} = \int d^3 R \int d^3 v \langle g^2 \rangle /
(F_0 V)$ (an overbar is used here to indicate a combined velocity space
average and an ensemble/volume average, to be distinguished from angle
brackets that indicate an ensemble average).  This uses the same
velocity weighting as found in the $W_{g0}$ component of the generalized
energy in Eq.~(\ref{W3d}).  After discretization, this becomes
$\overline{g^2} = 2 \tilde{\sum}_{\bf k} \sum_i m_i c_{i,i}({\bf k}) /
F_0({\bf v}_i)$. Using Eq.~(\ref{Gspec}) evaluated with the coefficients
given just before Eq.~(\ref{Dspec3d}) for the 3+2D case, and using
the same approximations as used just before Eq.~(\ref{phi3d}) (where
velocity summations are approximated by integrals assuming a
well-resolved velocity limit), one can show that
$$
\overline{g^2}
 = \frac{2}{\gamma} \left( N \tilde{\sum}_{\bf k} -
 \tilde{\sum}_{\bf k}
   \Gamma_0(k_\Perp^2) \right)
 \approx \frac{N N_k}{\gamma},
$$
for $N \gg 1$ (recall that $\sumkp$ is defined as a sum over the modes
in the upper half plane, so the total number of Fourier modes is $2
\sumkp = N_k$).
We thus find that the thermal noise spectrum in
Eq.~(5) of Ref.~\onlinecite{Nevins05} for $\delta f$ PIC codes (using a
weighted-particle Klimontovich representation for the distribution
function) is identical to the thermal spectrum calculated here for a
continuum code using a spectral representation for the distribution
function, with the identification of $1/\gamma = \overline{g^2}/(N N_k)$
in a continuum code with $\overline{w^2} / N_p$ in a PIC code. So the
total number of particles $N_p$ in the PIC code is
equivalent to $N N_k$, where $N$ is the number of velocity grid points
and $N_k$ is the number of Fourier modes in the continuum code, and
the mean squared particle weight $\overline{w^2}$ (which is called
$\langle w^2 \rangle$ in Ref.~\onlinecite{Nevins05}) is equivalent to
the continuum value of the mean-square particle
distribution function $\overline{g^2}$.  (From the PIC perspective,
this is consistent because the particle weights $w$ are equivalent to
$\delta f/F_0$, and in $\overline{w^2} = \sum_{i=1}^{N_p} \langle w_i^2
\rangle / N_p$, the marker particles have an $F_0$ distribution.)  This
equivalence between continuum and PIC thermal spectra is similar to
that found in 2-D hydrodynamics between Fourier-Galerkin and
point-vortex representations of the problem.  (The finite-size particles
used in most plasma PIC codes provides a kind of ultraviolet cutoff that
removes issues that could arise from point vortices or point particles
forming tightly-bound pairs.)

The thermal noise spectrum in PIC codes can be worked out using the
test-particle superposition principle, assuming that shielded test
particles can be considered independent and random.  The equivalence of
the PIC and continuum thermal spectra indicates that one can likewise
consider the random phase and amplitude of a Fourier-mode in $g$ at a
particular velocity (plus the plasma shielding of this mode) to be like
the random position and weight of a shielded test particle.

Since the thermal noise level $\langle {\phi}^2 \rangle =
2 \tilde{\sum}_{\bf k} \langle |\hat{\phi}_{\bf k}|^2 \rangle$ scales as
$\overline{g^2} / N$, it is important for both PIC and continuum codes
to either have enough particles or velocity grid points per spatial grid
point so that the noise does not get too large on the time scale of the
simulation, or to have enough small-scale dissipation to prevent the
particle weights or $\overline{g^2}$ from growing too large during the
simulation and causing a bottleneck problem\cite{FrischPRL08} or a
numerical diffusion problem.\cite{Nevins05}  (We have considered the
uniform plasma case in this paper where $\overline{g^2}$ is a constant
set by initial conditions, but in the case of turbulence driven by a
background gradient, $\overline{g^2}$ will increase in time due to an
entropy balance relation\cite{Krommes94,Nevins05} unless dissipation is
accounted for.)  Most continuum codes avoid such problems by employing
either physical collisions or numerical dissipation such as high order
upwinding or hyperdiffusion, though work on improved subgrid models
might be able to help optimize the performance by reducing resolution
requirements.  While $\delta f$ PIC simulations can formally work
correctly for a given simulation time period if enough particles are
used, eventually the noise can grow in time to become a problem.  PIC
codes can avoid this issue and/or reduce the particle resolution
requirements by employing weight-resetting methods,\cite{YChen08} which
essentially provide some numerical diffusion to limit the growth of the
weights.
% \nocite{*} % would cause all uncited items in the database to be included.
%\bibliography{gkAEaT0}% Produces the bibliography via BibTeX.

\begin{thebibliography}{10}
\bibitem{Notejzz}%
  \BibitemOpen
  \bibinfo {note} {JZZ is not responsible for all the computations and discussions in this section. Especially, JZZ believes there are subtleties in taking the continuous limit which can be illustrated with the direct formulation and calculation for the problem with continuous velocity, which is consistent with the present calculation with discretization of (or quantized) velocity, as shown in Ref. \onlinecite{Zhu20103}.}\BibitemShut{Stop}%
\end{thebibliography}
%Merlin.mbs v4.21 2009-07-09.
\providecommand{\noopsort}[1]{}\providecommand{\singleletter}[1]{#1}%

\end{document}